This is a preprint of the following chapter: Zielinski, B., Kahlmeyer, M., Böhm, S., Bonse, J., Florian, C. (2026). Surface Functionalization with Lasers: From Nanostructures to Applications. In: Bonse, J., Lasagni, A.F. (eds) Scaling of Laser Processing. Springer Series in Optical Sciences, vol 257. Springer, Cham. https://doi.org/10.1007/978-3-032-17133-7_27; reproduced with permission of Springer Nature Switzerland AG. The final authenticated version is available online at: https://link.springer.com/chapter/10.1007/978-3-032-17133-7_27

Chapter 27

# Surface Functionalization with Lasers: from Nanostructures to Applications

Bastian Zielinski[1,2], BastianZielinski@uni-kassel.de

Martin Kahlmeyer[3], m.kahlmeyer@uni-kassel.de

Stefan Böhm[3], s.boehm@uni-kassel.de

Jörn Bonse[4], joern.bonse@bam.de

Camilo Florian[1,2,*], camilo.florian@uni-kassel.de

1. Institut für Werkstofftechnik, Universität Kassel, Mönchebergstraße 3, 34125 Kassel, Germany
2. Institut für Physik, Universität Kassel, Heinrich-Plett-Straße 40, 34132 Kassel, Germany
3. Institut für Produktionstechnik und Logistik, Universität Kassel, Kurt-Wolters-Str. 3, 34125 Kassel, Germany
4. Bundesanstalt für Materialforschung und -prüfung (BAM), Unter den Eichen 87, 12205 Berlin, Germany

* Corresponding author: camilo.florian@uni-kassel.de

Pulsed laser sources stand out as remarkable and versatile tools in the growing interdisciplinary field of materials science, offering capabilities for tailoring and customizing materials to meet specific requirements. Their ability to induce different controlled material alterations allows enhancing material properties that might otherwise remain elusive. In essence, laser-induced material modifications encompass a spectrum of transformative processes, ranging from subtle topographical changes in the nano- and microscale, to intricate chemical modifications. Among the challenges to overcome are the creation of approaches that allow tuning the properties of light to optimize interactions on various types of solids, including metals, semiconductors, and dielectrics. Understanding these interactions is paramount towards efficiency increase of the processing techniques. This chapter embarks on a comprehensive exploration of the intricate dance between pulsed laser beams and their interaction with matter at the surface. The applications of laser-induced surface modifications are far-reaching and extend into widespread industrial fields. Examples of these include tribology, where the study of friction, wear, and lubrication is paramount. Besides that, the modified surfaces find applications in wetting phenomena, changing surface energy and contact angle control, enhanced welding processes, and optimizing molecular adhesion for stronger and durable interfaces. The ability to precisely engineer material surface properties through laser processing opens new frontiers in materials and surface sciences, offering innovative solutions and pushing the boundaries of what can be achieved in various industrial and technological domains.

# 1 Introduction

The following brief descriptions of intra- and inter-pulse feedback mechanisms, as well as laser modification on the nano and microscale, summarize the underlying processes and provide a classification of scales for the surface functionalization presented in Sect. 2. These aspects are explored in greater detail in Part I of this book. For example, Chap. 1 by Nolte et al. focuses on the interaction of matter with (multiple) laser pulses and Chap. 3 by Gräf et al. discusses the distinction between macro-, micro- and nanoscale processing with laser sources. Additional chapters relevant to these topics are highlighted below.

## *1.1 Intra- and Inter-Pulse Feedback Mechanisms*

When modifying materials with pulsed laser systems, the light-matter interaction leads to various feedback mechanisms that influence energy transfer and material modification [1]. Understanding these mechanisms and proposing laser irradiation approaches to control them towards their exploitation in applications become essential in different fields, as they dictate the efficiency of energy deposition and the nature of material modifications during laser processing. These mechanisms operate on different time scales modifying the material properties for a short period of time (transient changes) or permanently and can enable functionalities as discussed in this section. Considering the temporal frame where these changes occur, they can be categorized as intra-pulse and inter-pulse feedback mechanisms and organized as shown in Table 28.1.

**Intra-pulse Feedback Mechanisms** occur within the duration of a single laser pulse. During this brief period, typically ranging from femtoseconds to nanoseconds, several rapid processes can influence the laser-material interaction. The intense electric field of the laser pulse can ionize the material's surface, leading to the generation of free electrons and the formation of plasma. This plasma can absorb and reflect portions of the incoming laser energy, thereby modifying the energy deposition during the duration of the laser pulse. This effect is particularly significant in laser micromachining, where excessive plasma shielding can reduce the precision of material removal (see Chap. 1, Nolte et al.). Another important effect within the pulse interaction involves the excitation of electrons, which can cause rapid changes in the material's refractive index and absorption coefficient. Similarly, the irradiated material can undergo rapid structural phase transitions, such as thermal melting, which may, in turn, also abruptly alter the material's surface reflectivity on the picosecond timescale. These transient optical property changes affect how the remaining portion of the laser pulse propagates through and interacts with the material. Additionally, when the laser pulses provide a high intensity peak, nonlinear optical effects, such as multi-photon absorption, can also be triggered, modifying energy coupling into the material [2, 3].

**Inter-pulse Feedback Mechanisms** occur between consecutive laser pulses, over time frames significantly longer than the pulse duration. These mechanisms involve cumulative effects that result from the interaction of multiple pulses with the material, commonly referred to as incubation-effects. See Chap. 7, Lenzner et al., for a detailed discussion of these, some of which are summarized here. One of the most noticeable are topographical changes, occurring when repeated laser pulses lead to the gradual removal of material (ablation) or the reorganization of the surface structure, resulting in the formation of features such as ripples or grooves (see Chap. 10, Ancona et al., presenting the fundamentals of Laser-induced Periodic Surface Structures (LIPSS)). These surface modifications can alter the absorption and scattering of subsequent pulses, thus influencing the efficiency and outcome of laser processing. This effect is widely exploited in surface texturing for solar cells, where controlled structuring enhances light trapping efficiency. Given the current availability of high repetition rate femtosecond laser sources, thermal accumulation also plays a role, since depending on the repetition rate of the laser system and the thermal properties of the material, heat may accumulate between pulses. This thermal buildup can change the material's response to subsequent pulses, potentially lowering the ablation threshold or altering the nature of material removal. Additional changes may include chemical modifications that are influenced by surface temperature such as superficial oxidation or chemical permeation. Structural material changes, such as superficial amorphization, or polycrystalline solidification are usually driven by transient local heat gradients. This effect is particularly important in laser welding, where managing thermal accumulation and heating duration has a significant impact on the bond quality. Mechanical and thermal stress buildup can also lead to microcracks or material fatigue, influencing functionality of the final laser irradiated piece. This mechanism is leveraged in glass cutting with lasers, where controlled stress formation enables precise and predictable cutting results [4].

| | Mechanism | Effect | Application |
|---|---|---|---|
| Intra-pulse feedback | Plasma Shielding | Ionization creates plasma that absorbs or reflects part of the laser energy, reducing energy delivery to the material. | Superficial chemical interaction is enhanced. |
| | Transient Optical Property Changes | Rapid excitation of electrons alters the refractive index and absorption coefficient of the material during the pulse. | Marking resolution can be controlled in materials which structure is modified by laser light (i.e. phase change materials or photosensitive polymers). |
| | Nonlinear Absorption | High-intensity pulses induce multiphoton absorption, modifying the material interaction. | Transparent to light materials can be modified with high resolution. |
| | Thermal Conduction During the Pulse | Heat spreads within the material during the pulse, affecting temperature distribution and energy absorption. | Minimized thermal collateral damage with ultrashort laser pulses. |
| Inter-pulse feedback | Topographical Changes | Ablation modifies surface morphology, altering absorption and scattering of subsequent pulses. | Enhancement of effective total surface. Structural coloring. Light scattering. |
| | Heat Accumulation | Residual heat from previous pulses lowers the ablation threshold and changes material response. | Control of thermal buildup and solidification for consistent bond quality. |
| | Chemical and Structural Changes | Repeated pulses induce oxidation, phase transitions, or chemical modifications affecting laser-material interactions. | Surface functionalization for wetting, cell and bacterial adhesion. Increased hardness. Enhanced adhesion. |
| | Stress and Crack Formation | Mechanical and thermal stress buildup leads to microcracks or fatigue, influencing further laser interaction. | Laser-induced controlled cracking for precise material cuts and material fatigue control. |

Table 28.1 Intra- and inter-pulse feedback mechanisms for materials processing with pulsed laser systems. Adapted with permission from [1]

## *1.2 Laser Modifications on Different Scales: Nano vs Micro*

The irradiation of continuous areas with pulsed lasers typically begins with the formation of continuous lines, achieved by overlapping consecutive single laser pulses (see Figure 28.1a). Subsequently, when consecutive lines overlap, a continuous area is formed (see Figure 28.1b). Looking at the dimensions of the permanent material modification relative to the surface, we classify three categories, as summarized below and sketched in Figure 28.1c: First, the *Surface Level* topographic modification includes micro-sized changes in the form of craters, holes, trenches, including hierarchical structures that combine both micro- and nanofabricated surfaces, or purely nanometric features. Second, micrometric and nanometric *Debris* that can redeposit on the surface after material ablation. Third, modifications that start at the surface level and propagate inside the material up to a certain depth. Certainly, this also includes modifications at the surface, but the finally exploited functionality for applications lies in the bulk of the material.

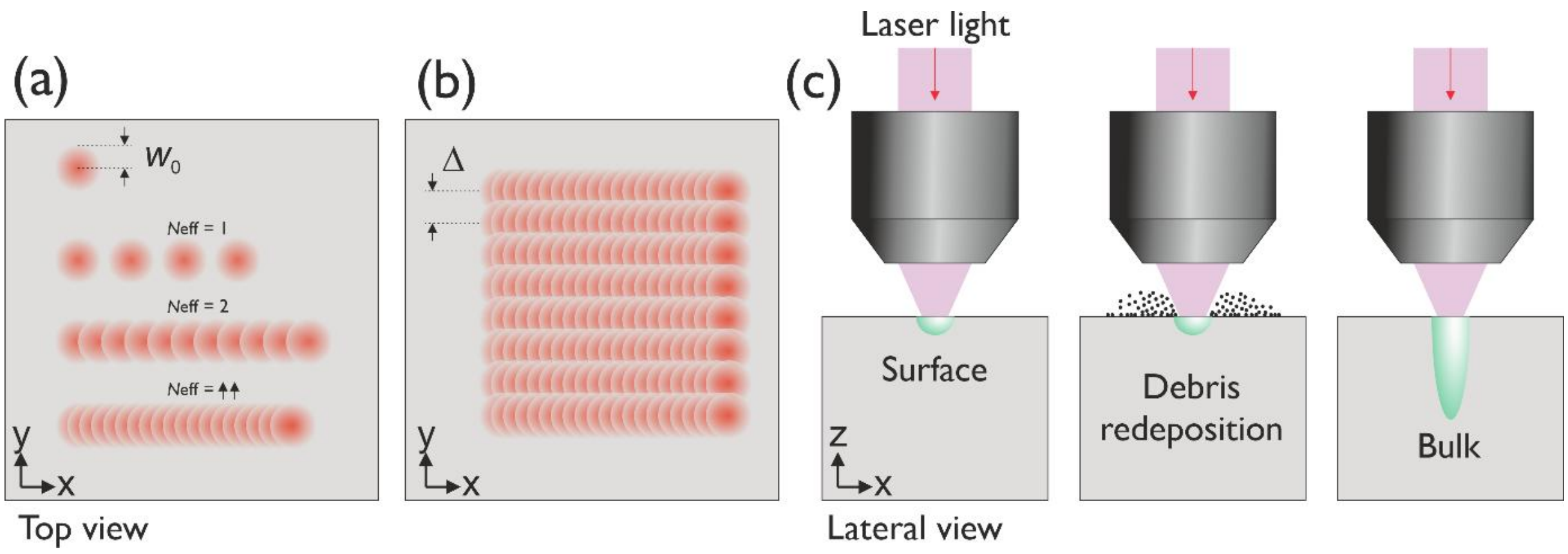

Figure 28.1 (**a**) Schematic of irradiation with pulsed laser systems where the beam radius at the materials surface is $w_0$. The interplay between the scanning speed ($v$) and the laser repetition rate ($\upsilon$) defines the number of effective pulses per spot unit area ($N_{eff}$), which increases, producing a continuous line. (**b**) Modified areas can be achieved by overlapping consecutive lines at given distance (Δ). (**c**) Laser modification of materials at different scales at the surface and in bulk material

### Modifications at the Surface Level

At the micrometer scale, pulsed lasers enable the fabrication of superficial structures that maintain a balance between high precision and reasonable processing speeds, which are, in some cases, compatible with industrial processing speeds. This scale is particularly relevant for applications in microfluidics, biomedical implants, and precision engineering components, where feature sizes in the range of a few microns to hundreds of microns are widely implemented [1]. One significant advantage of micrometer-scale processing with pulsed lasers is the ability to achieve structures with reduced collateral damage. Here, the highest finishing quality is achieved for the shortest pulse durations and wavelengths. For example, deep micro-drilling and cutting applications in metals, ceramics, and polymers benefit from the non-thermal nature of ultrashort pulse interactions, leading to clean edges and reduced recast layers. Such properties are crucial for applications in aerospace, medical devices, and semiconductor manufacturing. However, for most cutting, welding, and sintering applications with pulsed laser systems, longer pulses are widely used even at the cost of producing larger heat affected zones. In all cases, by controlling laser parameters - such as fluence, repetition rate, and scanning speed - precise micro-patterns can be generated to enhance performance of the modified surfaces.

A particularly interesting approach with pulsed laser processing is the creation of hierarchical structures, where microstructures are adorned by nanostructures [5]. These multi-scale surface modifications can enhance a material's functional properties by integrating the benefits of both scales. For example, in biomedical applications, hierarchical structures can improve cell adhesion and proliferation by mimicking natural textures found in biological systems [6]. Similarly, in tribological applications, microscale grooves combined with nanostructures can significantly reduce friction and wear [7] (see Sect. 3.1). The ability to create such hybrid structures opens new avenues for advanced material engineering, improving performance in diverse fields such as optics, energy storage, and sensor technology.

At the nanometer scale, pulsed lasers offer generally precision fabrication. Sub-wavelength feature sizes can be produced either by direct laser writing, in some materials through non-linear absorption processes, or by exploiting plasmonic behaviors at the surface, to produce modifications known as laser-induced periodic surface structures (LIPSS) (see Chap. 10, Ancona et al.). For instance, multi-photon absorption enables material processing of arbitrary patterns below the diffraction limit of the laser wavelength, allowing for nanoscale resolution in laser-based lithography and patterning by direct writing, even when combining multiple materials [8] (see Chap. 27, Mao et al.). This phenomenon is instrumental in developing advanced optoelectronic components and high-resolution data storage solutions. On the other hand, for ultrashort laser pulses, LIPSS are frequently observed when processing materials with multiple laser pulses. These self-organized patterns are often formed due to the interaction between the incident laser light and surface plasmon waves, and can be leveraged to tailor surface functionalities, such as wettability, reflectivity, and biocompatibility [9]. While they can reach periodicities an order of magnitude below the laser wavelength, their feature size is independent from the dimensions of the laser focus. The laser power and repetition rate define how large an area is patterned at once.

The choice and balance between nano- and microscale processing depends on the intended application and material characteristics. Microscale processing provides a good balance between precision and efficiency but may not achieve the ultra-fine resolution required for certain high-end applications. In contrast, nanoscale processing by direct writing offers extreme precision, but it is often limited by lower throughput and increased complexity in controlling non-linear absorption phenomena. In cases where simple periodic nano-patterns are useful, structuring with LIPSS can be an efficient strategy. Furthermore, as feature sizes shrink, challenges such as debris redeposition, surface roughness control, and uniformity become more pronounced. Advances in beam shaping techniques, adaptive optics, and real-time feedback mechanisms are being developed to overcome these limitations and improve process reliability on industrial scales, see Chap. 20 by Bonse et al.

It should be noted that the embossing of laser-generated nano- and microscale surface structures into (softer) materials enables production and application on a large scale. Chap. 34 by Hennig et al. discusses the challenge of creating structured surfaces on large print rollers and presents relevant applications.

**Debris Redeposition**

When the laser fluence is sufficiently high at the surface of the material of interest, material ablation occurs, leading to a rapid removal of material and the formation of debris, which can redeposit onto the surface [10]. The nature of the debris depends on several factors, including the material type, laser parameters (such as pulse duration, wavelength, and fluence), and surrounding environmental conditions. Debris generated during ablation can take various forms, including nanoparticles, micro-dust, molten droplets, and vaporized material. Mechanisms of debris formation include the direct ejection of solid particles, phase explosion due to rapid heating of the irradiated material towards the thermodynamic critical point, and plasma-induced material breakdown. In some cases, vaporized material may recondense into nanoparticles that settle back onto the surface, forming a redeposition layer. In contrast, larger molten droplets can be expelled due to recoil pressure and cooling-induced fragmentation. It is worth noting that this effect can also be exploited to produce useful nanoparticles for other applications, as discussed in Chap. 29 by Bulgakov et al.

In ultrashort pulsed laser processing (femtosecond or picosecond range), ablation is often characterized by minimal heat diffusion, resulting in fine, well-defined debris with reduced thermal effects [11]. Still, ejected molten material, in the form of nanoparticles or micro-dust, may adhere to the surface and form a layer of debris. At the nanometer scale, redeposited nanoparticles can alter optical or electronic properties, especially when embedded into dielectric passivation oxides while at the micrometer scale, micro-dust can lead to surface contamination and increased roughness.

For longer pulse durations, such as nanosecond or microsecond pulses, heat diffusion plays a more significant role in the ablation process. As a result, material removal is often accompanied by the formation of larger molten droplets and increased recast layers. These longer pulses cause more extensive thermal effects, leading to larger heat-affected zones (HAZ) surrounding the irradiated surface spot and rougher surfaces due to re-solidified material. Additionally, vaporized material may condense into larger particulates, contributing to microdust accumulation and further altering the final surface quality. The combination of these processes can contribute to the creation of nanostructures and even hierarchical structures that influence surface quality and overall surface performance.

In the context of surface functionalization for applications, structures created by debris can be beneficial or detrimental, as will be discussed in Sect. 3.2 (wettability) and Sect. 3.4 (adhesion). In case the effects by debris are unwanted, strategies such as gas-assisted laser processing, optimized scanning strategies, and post-processing cleaning techniques can mitigate debris accumulation, ensuring high-quality surface structuring.

**Modifications in the Bulk**

Laser light can penetrate materials transparent to their wavelength and create modifications inside the bulk material, such as the creation of voids or laser-induced chemical and structural changes. Regarding the functionalization of surfaces discussed here, *modification in the bulk* means that the modification starts at the surface and can extend to a certain depth depending on the material's optical properties and the laser parameters used, including pulse duration, optical penetration depth, laser fluence, and repetition rate. In metals, deeper modifications are largely driven by heat accumulation, either from longer pulse durations or high repetition rates. In such cases, when laser irradiation occurs in reactive gas atmospheres, additional chemical bonds can form, altering both the mechanical and chemical properties within a specific depth. This effect is extensively studied in laser peening, a process used to enhance fatigue resistance and mechanical strength in metals by inducing compressive stress through controlled irradiation [12].

Beyond laser peening, bulk modifications by pulsed lasers play a crucial role in other material processing applications. In welding, the application of pulsed lasers with tailored heat input can effectively influence microstructures and precipitation, even in deep layers of the weld (see Sect. 3.3). This leads to significant improvements in material properties, such as strength, corrosion resistance, and overall weld performance [13]. Similarly, surface preparation for adhesion benefits from controlled laser-induced roughening and chemical activation, enhancing the bonding strength between materials [14] (see Sect. 3.4). Additionally, particle and ion implantation, as well as laser hardening techniques, leverage in-depth modifications to improve wear resistance and mechanical durability in industrial components [15, 16] (see Sect. 3.1). The ability to tailor modifications at different depths opens new possibilities for advanced material processing, where bulk properties can be fine-tuned for specific applications. By adjusting laser parameters and environmental conditions, manufacturers can optimize the mechanical, chemical, and structural characteristics of a material, enhancing performance across industries such as aerospace, automotive, and biomedical engineering.

# 2 Relevance of Topographical and Chemical Changes

In this section, we explore the critical role of both topographical and chemical modifications in the functionalization of materials for various applications. We begin by examining how surface structuring enables structural coloring, focusing on recent laser irradiation techniques that create periodic surface features capable of diffracting light. As we will demonstrate, even in applications that primarily rely on surface morphology, laser-induced chemical changes often play an indirect but decisive role in enhancing or even driving functionality. In the second part of this section, we shift our focus to cases where chemical modifications are the primary mechanism behind the material's functional transformation. Here, we will review key experimental demonstrations where laser-induced alterations in composition, oxidation states, or phase transitions directly govern the application's effectiveness, highlighting the interplay between structural and chemical changes in laser processing.

## *2.1 Surface Micro- and Nano-Texturing of Materials with Pulsed Laser Sources*

The study of laser-induced structural colors began with the discovery of LIPSS in 1965 [17]. These periodic structures have been labeled with different names due to the wide variety of morphologies and periodicities, which range from a few nanometers to several microns. These structures generally form when a material surface is irradiated with short and ultrashort pulses of laser light, creating wavelength- and sub-wavelength-sized periodic structures. When these structures interact with incident light, they can act as a diffraction grating and thus produce vibrant colors under specific viewing angles, as will be discussed here. Additionally, LIPSS textured surfaces also find applications in optimizing tribological performance, controlling wettability, and adhesion, as it will be discussed in Sect. 3.

Early studies primarily focused on understanding the formation mechanisms of these structures, which were influenced by a handful of parameters, including the laser wavelength, pulse duration, angle of incidence, and polarization [18]. The development and broad availability of femtosecond laser systems in the 1990s promoted significant advancements, as the ultrashort laser pulses enabled highly controlled energy deposition, minimizing thermal effects and allowing precise nanoscale modifications [19]. These breakthroughs paved the way for the controlled generation of structural colors on various materials. As research progressed, scientists explored the use of femtosecond lasers to fabricate tailored nanostructures for full-spectrum color generation. By adjusting laser parameters such as fluence, scanning speed, and pulse overlap, a wide range of colors could be achieved on surfaces like titanium [20] and stainless steel [21]. These colors were found to be highly stable, environmentally resistant, and angle-dependent due to the interaction between the surface topography and incident light.

Notably, in recent years, advances in illumination strategies have further expanded the capabilities of laser-induced coloration. Zhang et al. have reported a novel approach in which multiple laser writing cycles with varying polarization directions produced vibrant and stable structural colors on steel [22]. These structures were fabricated using an average laser power of 120 mW, scanning speed of 20 mm/s, and laser pulse energy of 4.8 µJ ($\tau = 50$ ps, $\lambda = 1030$ nm, $\upsilon = 25$ kHz, $w_0 = \sim 30$ µm). The periodic structures formed in patches with different orientations and did not completely overlap. Therefore, two orientations were imprinted in the material at the same time, which under illumination with white light diffracts the light in different directions. Figure 28.2 includes different areas irradiated with the double scan approach, showing the surface topography and the resulting functional color under different angles (more angles are available in [22]). This innovation allowed for full-color marking with enhanced viewing-angle independence, significantly improving the practicality of laser-induced colors for industrial applications.

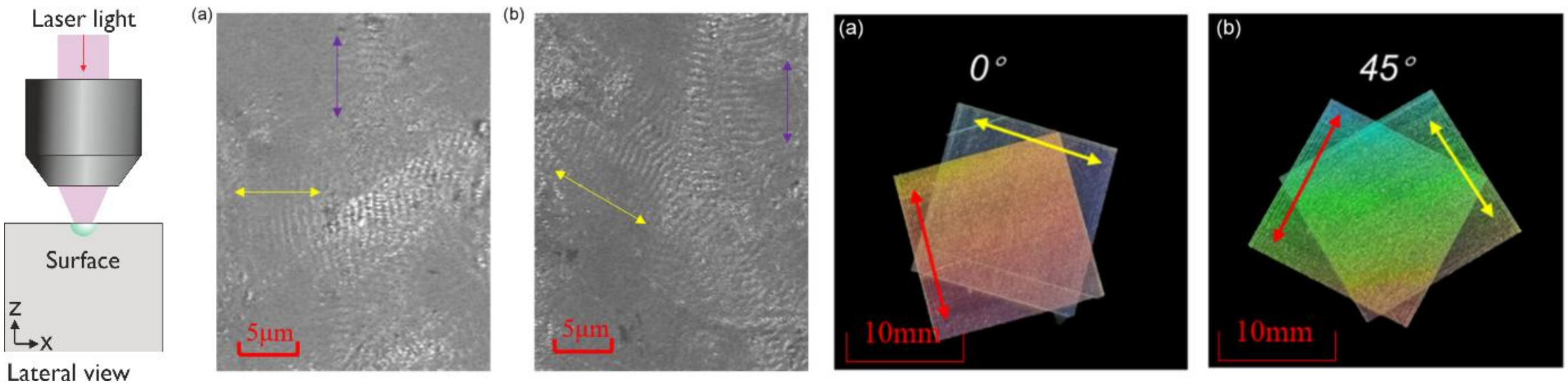

Figure 28.2 The left part shows top-view scanning electron micrographs (SEM) of the double-laser-scanned areas, each processed with a different polarization orientation as indicated by the yellow and purple double arrows. The right part shows the resulting coloring of one sample under different observation angles and how the colors of the two orientations are superimposed. Reprinted (adapted) with permission from Zhang et al. [22] under Creative Commons CC BY license, Advanced Photonics Research published by Wiley-VCH GmbH

Recent advancements in laser-induced structural coloration have significantly enhanced the efficiency, resolution, and functionality of this technology, paving the way for diverse industrial applications. For instance, Geng et al. [23] recently demonstrated a method capable of producing structural colors at a rate of 1.4 cm²/s, achieving a printing resolution exceeding 25,000 dots per inch (dpi). This technique utilizes ultrafast laser pulses to induce photothermal oxidation on ultrathin hybrid films, resulting in durable and angle-robust colors suitable for high-throughput industrial applications ($\tau = 10$ ps, $\lambda = 1030$ nm, $5$ kHz $< \upsilon < 75$ kHz, $w_0 = \sim 120$ µm). These advancements underscore a significant trend toward integrating laser-induced structural coloration with functional applications, particularly in security and data encryption, as they demonstrate robustness and high performance, as displayed in Figure 28.3 (left).

In parallel, advancements in ultrafast laser processing have led to unprecedented resolutions in color marking and information encryption. Lapidas et al. reported a direct femtosecond laser printing approach that achieves surface coloring and plasmonic information encryption at an extraordinary resolution of up to 50,000 dpi [24]. This method leverages the precise control of laser-induced plasmonic nanostructures to modulate surface reflectivity, enabling high-density data encoding within colored images. Such high-resolution capabilities open new avenues for applications in anti-counterfeiting measures, high-density data storage, and secure information encryption ($\tau$ = 200 fs, $\lambda$ = 515 nm, $w_0$ = ~700 nm, 1 mm/s < $V$ < 400 mm/s). Depending on the laser fluence used, the multilayered system develops different structures including nano bumps, holed bumps and complete ablation, enabling control over the morphological sizes of the modifications giving place to different plasmonic behaviors.

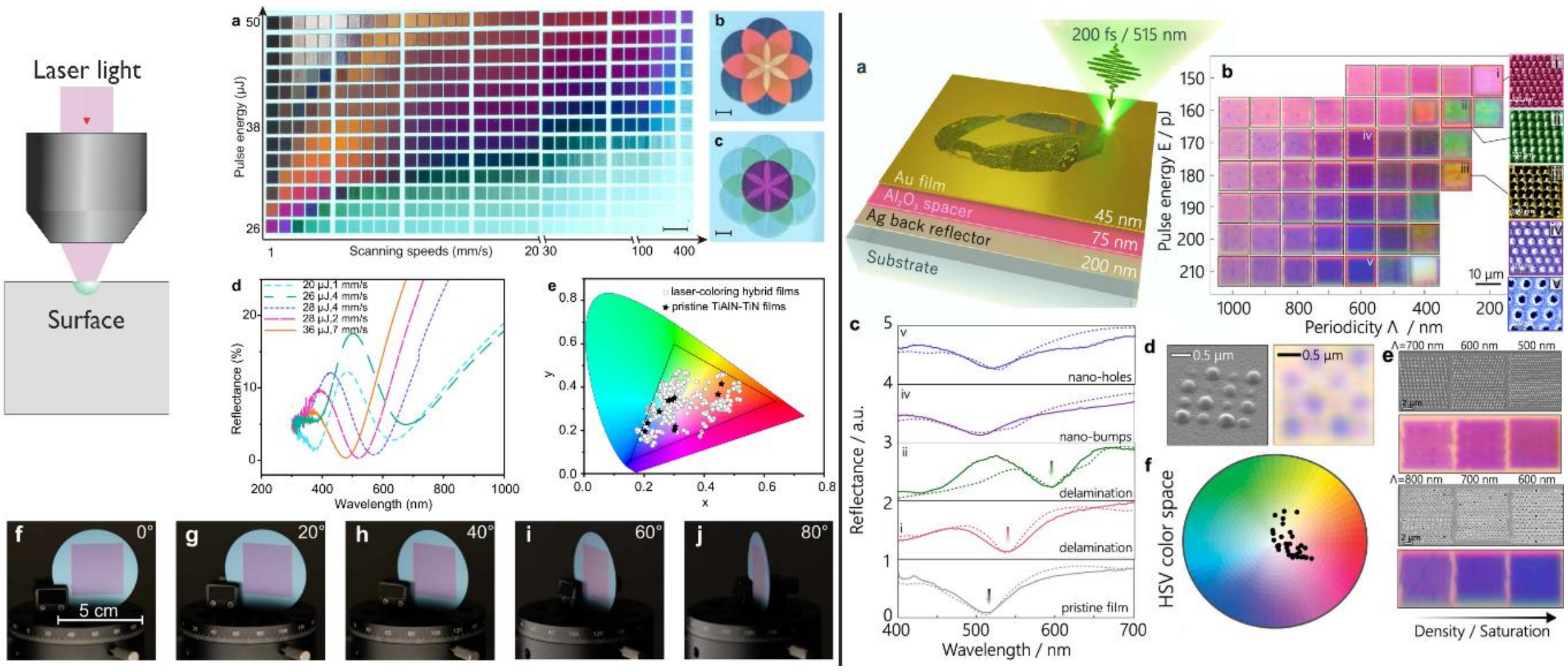

Figure 28.3 On the left: (**a**) Diffraction color characterization presented as a matrix palette of the structures produced at different scanning speeds and fluences. (**b**) and (**c**) include the irradiation of seven overlapped circle-shaped areas that demonstrate re-writability of colors. (**d**) Reflectance spectra of several representative areas. (**e**) Chromaticity diagram. (**f-j**) Photographs of a large area processed with the structures, displaying that color performance does not depend on the viewing angle. Reprinted (adapted) with permission from Lapidas et al. [24] under open access license, Copyright © 2024 American Chemical Society. On the right: (**a**) Sketch that includes the multilayered substrate where the irradiation took place. (**b**) Color palette induced by the structures (10 × 10 $\mu m^2$) fabricated by varying the incident pulse energy E and nanostructure periodicity Λ (inset SEM images). (**c**) Calculated (dashed) and measured (solid) curves of reflectance of some representative areas. (**d**) SEM and bright-field images of the induced structures. (**e**) Examples of colors from nanobumps (top panels) and nanoholes (bottom panels). (**f**) Reflectance spectra of the colors shown in (**b**) represented as coordinates in the HSV color space. Reprinted with permission from Geng et al. [23] under Creative Commons CC BY license, Springer Nature

The ability to produce high-resolution, durable, and customizable colors on various substrates without the need for inks or dyes not only offers environmental benefits but also enhances the versatility and security of printed materials. As laser processing technologies continue to evolve, we can anticipate further improvements in processing speeds, resolution, and functional integration, solidifying the role of laser-induced structural colors in modern manufacturing and information security sectors.

While the discussion here focuses on LIPSS, other types of laser-generated features for decorative and security applications, like plasmonic nanoparticles and interference patterns, exist and are presented in Chap. 36 by Killaire et al.

Another type of nano- and microstructures that change the surface topography are site-selective, high aspect-ratio pillar-like nano-features, which are created by a single laser pulse and protrude above the original surface plane. Such structures have been reported as both *modifications at surface level* (see Figure 28.4) and as *modifications in bulk* (see Figure 28.5) – provided that specific boundary conditions (laser beam focusing conditions and involved materials/stratigraphy) are fulfilled (see below). Under these optimized conditions, material expulsion from the surface can be triggered and rapidly redistributes sample material during the transient relaxation stage, following the ultrashort pulsed laser excitation and localized melting of the sample. As a result, anisotropic surface structures with sub-micrometer diameters and heights ranging between a few and several tens of micrometers can be manufactured at desired sample sites and even extended from a single surface feature into two-dimensional arrays. In the following two figures, two examples of such processes are provided for "film nanoexpulsion", as well as for "bulk nanoexpulsion", respectively.

Figure 28.4 compiles a sequence of tilted-view SEM images of a 60 nm thick gold film on a fused silica substrate after irradiation by single ultrashort laser pulses ($\tau = 30$ fs, $\lambda = 800$ nm) of different energies (fluences) that were focused by microscope objective (NA = 0.5) directly onto the film surface [25]. In all micrographs a pedestal bump is seen after the laser exposure at the film surface, featuring heights up to ~ 800 nm. For laser pulse energies exceeding 7 to 8 nJ, an additional central protruding "nanojet-feature" has solidified at the center of the bump. It exhibits heights up to 1.5 µm and a laser pulse energy dependent aspect ratio up to ~5 in this case.

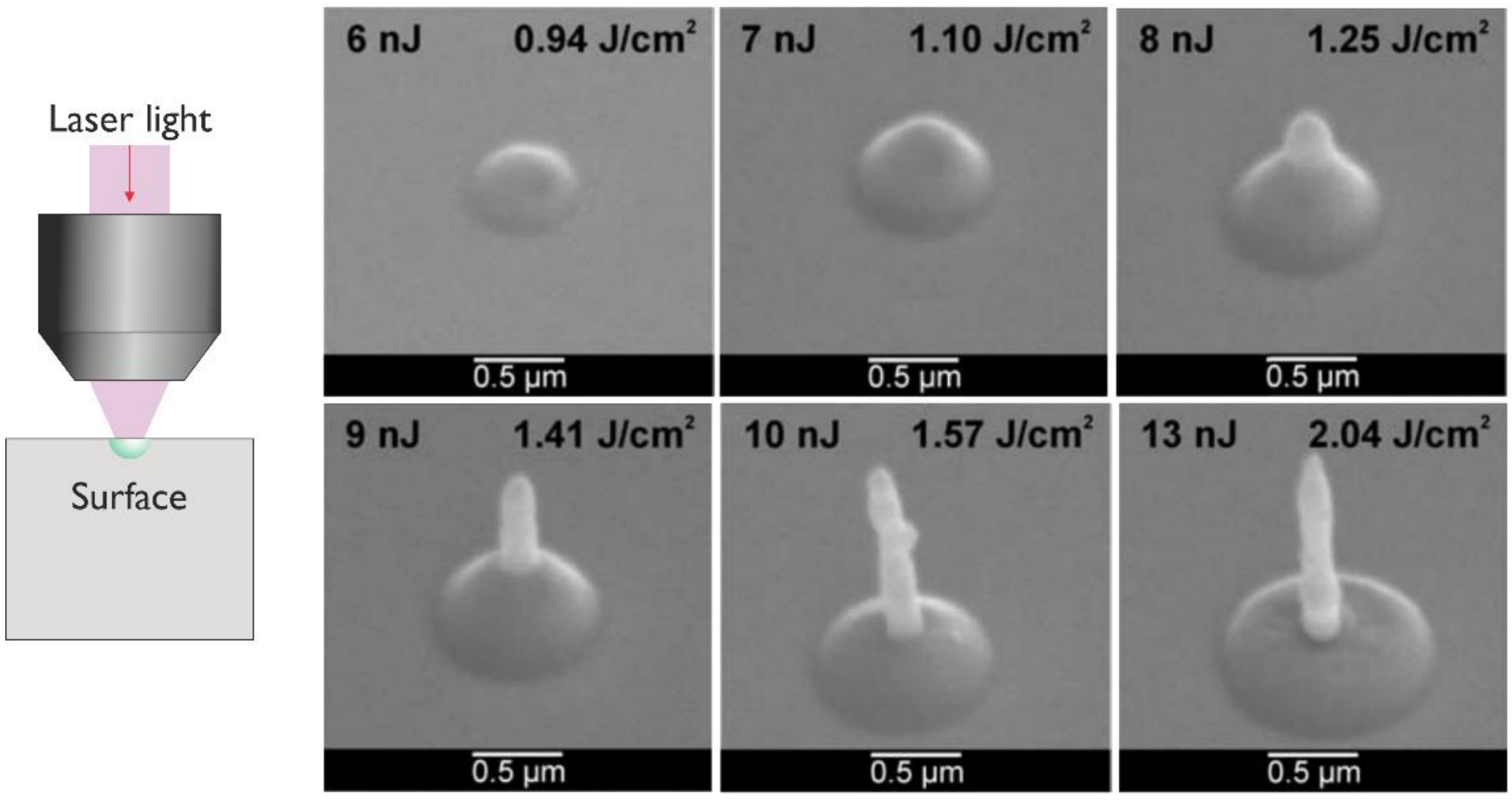

Figure 28.4 Side-view SEM imaging of sub-wavelength surface structures generated on a 60 nm thick gold film (deposited on a fused silica substrate) after irradiation by single, tightly focused Ti:sapphire fs-laser pulses (30 fs, 800 nm) at varying pulse energies between 6 and 13 nJ. Reprinted with permission from Korte et al. [25], Copyright (2004) Springer-Verlag

The topic of laser-induced frozen nanojets gained attention around 2004 after a Japanese and a German research group had independently reported the phenomenon [25, 26]. These publications then triggered further theoretical and experimental research that involved thermoelastic continuum modeling [27], atomistic molecular dynamics (MD) simulations [28, 29], and even time-resolved optical measurements [30]. The numerical modeling studies showed that the bump pedestal is caused by thermoelastic stress that triggers a delamination of the gold film from the dielectric substrate. The nanojet itself forms from the molten gold film and its hydrodynamic melt displacement and final ejection through inertia. The height of the transient jet is limited by surface tension effects through the Rayleigh-Plateau instability of the molten material [27, 29, 30]. Side-view shadowgraphic measurements revealed that the entire melting and re-solidification dynamics are completed within ~ 250 ns [30]. From these experiments, the following physical scenario can be drawn: About 5 ns after the absorption of the fs-laser pulse at the surface, the gold film delaminates, bulges and forms a bump of ~ 4.5 µm in diameter and of ~ 2.4 µm in height. At a temporal delay of ~ 35 ns, the bulge structure collapses, and a jet of molten material starts to be expelled. While the bump height further decreases, the central nanojet of molten material can grow to a height of ~ 4 µm, while continuously expanding until a delay time of ~ 125 ns. After ~ 150 ns, the nanojet partially detaches from the bump. Caused by the Rayleigh-Plateau instability, the nanojet disintegrates into two components, i.e., a small spherical droplet at its tip and a larger elongated droplet. After the detachment of the film material, a very thin sub-micrometric pillar remains at the bump before solidification sets in. For more details on the laser processing of thin films by ultrashort laser pulses the reader is referred to the review article [11].

Another impressive example of *modification in the bulk* in the form of nanostructure expulsion was recently reported by Belloni et al. [31]. The authors demonstrated the fabrication of high-aspect-ratio, nanopillars with a diameter of ~800 nm and heights up to 15 µm at the surface of sapphire upon single laser pulse irradiation ($\tau = 115$ fs, $\lambda = 800$ nm). For that, the authors employed diffraction-free, first-order Bessel beams with radial or azimuthal polarization and carefully controlled the laser pulse energy. In this configuration, the high-intensity laser-matter interaction locally melts the material in the doughnut-like high-intensity regions and generates a tubular-shaped, high-pressure field beneath the sample interface that leads to the rapid expulsion of material across the surface. This concept can be applied either at the front or the rear surface of the sapphire samples (see Figure 28.5a).

Depending on the laser fluence (controlled via the laser pulse energy) different nanopillar morphologies were observed with three main regimes (see Figure 28.5b). For the lowest fluence regime, cylindrical straight nanopillars are formed by the expulsive translation of solid material from the bulk. In the second fluence regime, smooth nanopillars are formed by the extrusion of highly viscous molten material. In the third fluence regime, several shapes of wavy nanopillars were observed and were explained in the framework of the Rayleigh-Plateau capillary instability theory of hydrodynamic jets of low viscosity liquid of the transiently molten sapphire.

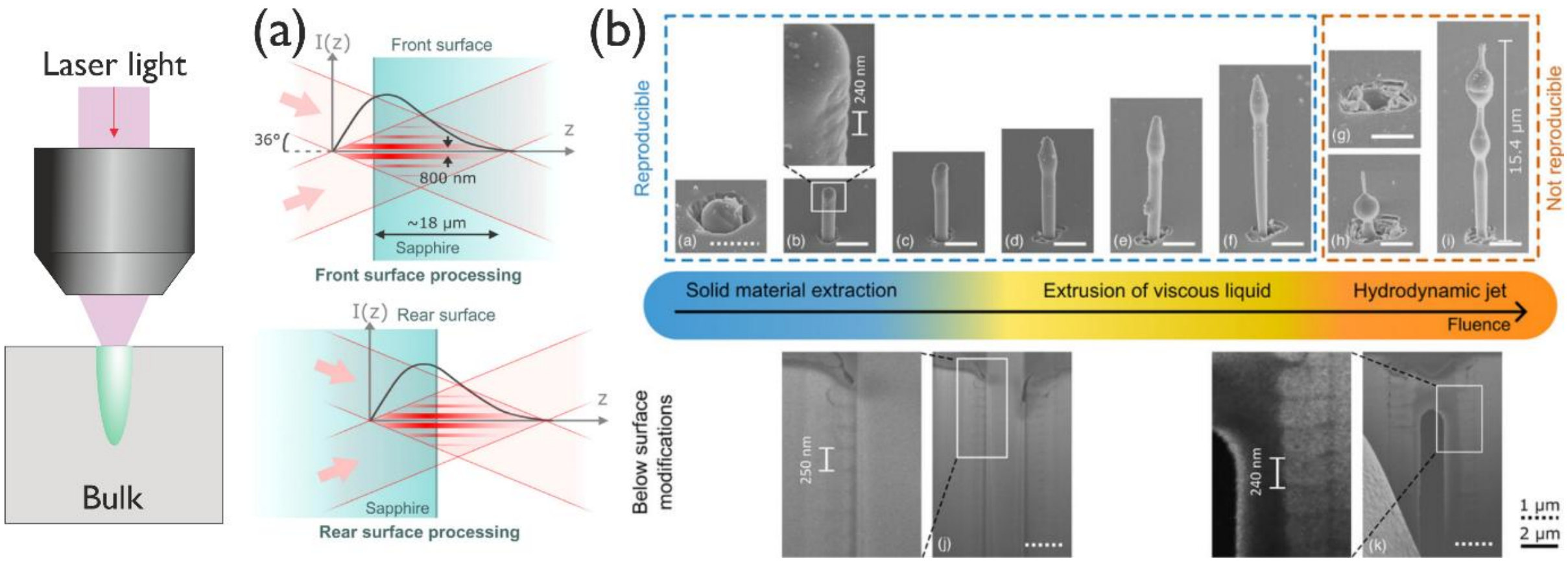

Figure 28.5 (**a**) Schematic of the laser irradiation strategy for inducing the expulsion of nanopillars from the bulk of sapphire. An ultrafast, first-order Bessel beam is focused through the sample front or exit surface. The resulting intensity pattern in the axially extended focal region is shaped as a hollow cylinder surrounded by several rings. The black solid line indicates the intensity of the cylinder along

the propagation direction. In both configurations, the intensity is kept higher inside the bulk than at the surface. (**b**) Classification of the nanopillar morphologies into three different regimes of fluences in the main Bessel ring. For the first and the second regime, the nanopillars are reproducibly formed. For higher fluence, the nanopillars cannot be generated in a reproducible manner. At the bottom part of the figure, electron micrographs of the corresponding subsurface modifications are displayed. Reprinted with permission from Belloni et al. [31] under Creative Commons CC-BY-NC-ND license, Copyright © 2024 Laser & Photonics Reviews, by permission from John Wiley and Sons

The Rayleigh-Plateau theoretical framework successfully predicted the period of the thickness-oscillations along the nanopillar axis, as well as the value of the maximum nanopillar height. Increased laser pulse durations into the picosecond range reduce spatial gradients of the temperature field, thus suppressing the first regime in which straight nanopillars of solid material can be ejected from the surface.

## 2.2 Chemical and Structural Changes induced by Pulsed Laser Sources

Pulsed laser systems play a crucial role in inducing chemical changes across various applications. In this subsection, we center on two applications: welding and surface functionalization. One of the most significant transformations induced by pulsed laser welding is the alteration of crystal sizes within the material. The rapid heating and cooling cycles characteristic of pulsed laser irradiation influence grain growth, phase transitions, and precipitation behaviors, directly affecting the mechanical properties of the irradiated materials. In general, the rapid thermal cycling associated with pulsed laser welding leads to refined microstructures due to the suppression of excessive grain coarsening. The high cooling rates prevent the extensive diffusion of elements, resulting in the formation of finer grains compared to conventional welding methods [32]. Additionally, the precise energy deposition of pulsed lasers allows localized control over the thermal input, which can be optimized to balance grain growth and precipitation kinetics in metals [33] and even in transparent materials [34]. A notable example of this effect is observed in the welding of AISI 430 ferritic stainless steel [35]. In this study, it was demonstrated how the pulse duration in the millisecond regime strongly influences the grain structure evolution. The study found that the application of tailored pulse profiles effectively controlled grain coarsening, promoting a more homogeneous microstructure while also affecting the distribution of precipitates within the weld zone. This level of control is essential for maintaining desirable mechanical properties such as toughness and corrosion resistance in stainless steel components [35].

Recently, as illustrated in Figure 28.6, a detailed study on the grain structure and formation in the laser irradiated weld region using ferritic stainless steel was presented by Sommer et al. [13]. Here, the effects of the pulse duration used during the laser welding process were investigated, paying particular attention to the grain growth and precipitation kinetics with AISI 4.0 FSS stainless steel ($\tau$ = 1, 3 and 5 ms, $\lambda$ = 1064 nm, $\upsilon$ = 5 Hz, $w_0$ = 33 µm, $V$ = 15 mm/min). Although the scanning speed in this case may be too slow for practical industrial applications, increasing the repetition rate to current industrial standards operating in the kHz regime may lead to an overall processing speed increase of a thousand-fold, allowing linear fabrication at m/min. Three cross-sections are included for the different pulse durations implemented. The false-color gradient allows a quick visualization of the grains that are bigger than those in the material bulk. When the pulse duration is the largest studied, the grains are bigger. Temperature gradients and solidification rates played an important role in the formation of the welded region, producing overall larger grain sizes oriented preferentially perpendicular to the laser impinging direction. Analysis of the weld metal allows the characterization of crack formation and intergranular corrosion, which increases with the grain size. Compared to continuous-wave laser welding, intergranular corrosion susceptibility is significantly reduced for very short pulse durations. These results suggest that pulsed laser beam welding with temporal pulse shaping allows for direct control of heat input, enabling tailored grain growth and precipitation formation, thereby improving corrosion resistance.

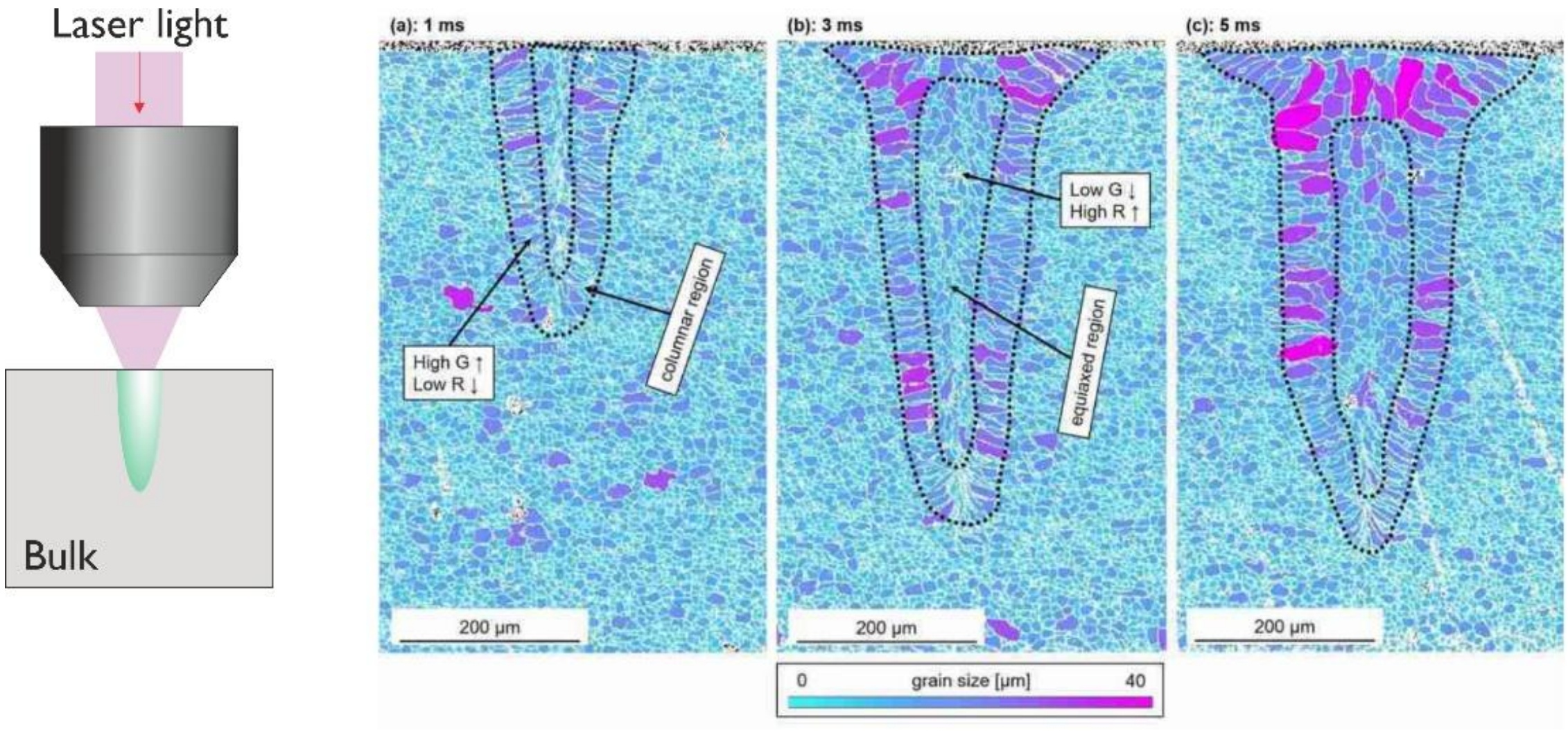

Figure 28.6 Lateral cross-sections of the laser-modified regions, extending from the surface at the top to the bulk material at the bottom. (**a**)–(**c**) Analysis of grain size evolution at various pulse durations, showing the transition from columnar to equiaxed. Dashed lines represent interfaces of distinctive microstructural features. Regions of high/low temperature gradient ($G$) and solidification rate ($R$) are also indicated. The molten region visibly produced larger grain sizes as the laser pulse duration ($\tau = 1$, 3, or 5 ms) increases. Reprinted with permission from Sommer et al. [13] under Creative Commons CC-BY license, Copyright © 2023 Springer Nature

Beyond microstructural refinement, pulsed laser irradiation can also induce significant chemical changes by facilitating the incorporation of foreign species into the material at the surface and bulk level for the functionalization of surfaces. This phenomenon occurs through various mechanisms, including laser-induced plasma interactions, ablation-driven deposition, and laser-assisted diffusion of atmospheric or pre-deposited elements into the substrate. The highly energetic nature of pulsed lasers can create conditions for oxidation, nitridation, and even carbide formation, depending on the surrounding environment and material composition. Chemical interactions driven by laser irradiation are also evident in nanomaterial synthesis where complex chemical reactions can be triggered by the laser irradiation [36, 37]. A specific case study highlighting the impact of femtosecond laser irradiation on the chemical incorporation of surfaces is included in Figure 28.7. The study by Florian et al. [38] demonstrated that ultrafast laser pulses not only restructure the titanium alloy surface but can also promote the diffusion of oxygen in depth to extents ranging from few nanometers up to several micrometers, leading to the formation of oxide and nitride layers. These chemical modifications enhance properties such as hardness, corrosion resistance, and biocompatibility, making them highly relevant for biomedical implants and aerospace applications.

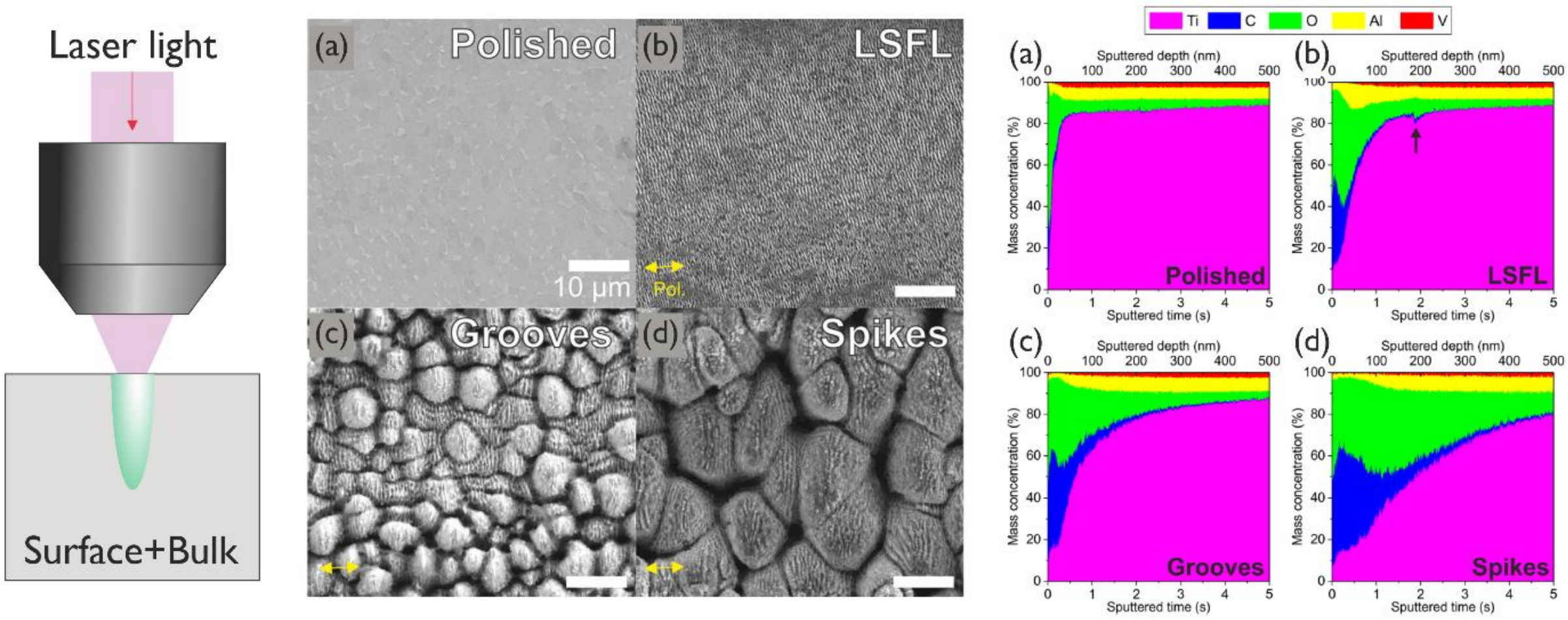

Figure 28.7 On the left: Top-view SEM images of the titanium alloy (Ti6Al4V) surface before and after fs-laser processing. (**a**) Micrograph of the polished sample surface. Laser structured areas with (**b**) LSFL, (**c**) Grooves, and (**d**) hierarchical Spikes. The yellow double arrow indicates the polarization direction of the laser beam. On the right: Depth profiles obtained from glow-discharge optical emission spectroscopy (GD-OES) showing the mass concentration of the different compositional elements of the alloy at different sputtered depths corresponding to the structures shown on the left pane. The vertical arrow in (**b**) marks a measurement artifact arising from plasma instability during the measurement. Reprinted (adapted) with permission from Florian et al. [38] under Creative Commons CC-BY 4.0 license, Copyright © 2020 MDPI Switzerland

Overall, pulsed laser irradiation serves as a powerful tool for inducing material transformations at surfaces, either through induced grain growth during welding or through the targeted incorporation of chemical species. By leveraging these processes, researchers and engineers can develop materials with tailored properties for a wide range of industrial and technological applications, as presented in the following Sect. 3 as well as a variety of chapters in Part III of this book.

# 3 Applications of Surface Treatments with Pulsed Laser Sources

In this section, we will survey selected publications on applications that incorporate the use of pulsed laser systems on applications for tribological performance, wetting on industrially relevant materials, welding and adhesion applications.

## *3.1 Optimization of Tribological Performance*

One of the first experimental demonstrations on the use of LIPSS for tribological applications was published around 1999 [39]. As discussed in Sect. 2, the laser-based *modification of a surface* can change its topography, its chemical properties, and the surface interaction with liquids. All these factors can contribute to tribological performance, which can be characterized by the coefficient of friction (COF, see Background Information) and the resulting wear during the relative motion of two bodies in contact with each other. This section will present examples where surface structuring improves friction and reduces wear, how this relates to nanostructures (LIPSS) and microstructures on different materials, and which altered properties of the treated surface are responsible. We will see that tribological performance differs, depending on materials, structures, and interaction.

Friction and wear are relevant for nearly all moving parts in current industrial machinery: friction translates energy, that's input to create motion, into heat, reducing energetic efficiency. The effects of friction can include deformation and removal of material from the surface, which contribute to the wear-down of a technical element. Wear can decrease efficiency and lead to failure and the need for replacement. This increases costs and reduces sustainability.

The COF is defined as the ratio between the force needed to overcome the friction between the two bodies and the force pressing them together. When comparing COF values, it has to be noted that there are different (standardized) tests to measure the COF with different geometries for addressing different tribological conditions, including the contact pressure and relative velocity of the involved bodies, lubrication, etc., such as Ball-on-Disk, Ring-on-Disk, or Pin-on-Disk tests, and test materials [40]. Therefore, research typically presents a comparison between untreated and laser-processed surfaces, where the relative change of COF and wear is evaluated.

In comparative studies of different steel and titanium alloys, the improvement of the COF with LIPSS-type nanostructures was significantly different [41, 42]. The surfaces were treated with linearly polarized laser pulses ($\tau$ = 30 fs, $\lambda$ = 790 nm, $\upsilon$ = 1 kHz) focused down to a Gaussian beam profile with radius $w_0$ ($1/e^2$) ≈ 140 µm. By scanning with a laser peak fluence of $\phi_0 \approx 0.11$ J/cm² and an effective number of pulses per spot of $N_{\mathrm{eff}} \approx 56$, LIPSS with periods in the range of 500 - 600 nm were created. Characterization of the COF was performed with reciprocating sliding tribological tests (RSTTs) against a polished and hardened 100Cr6 steel ball with diameter of 10 mm, normal force $F_N$ = 1.0 N in 1,000 cycles at 1 Hz.

As lubricants, paraffin oil or a fully formulated engine oil with additives (Castrol VP-1) were used, leading to the first observation that for paraffin oil there was no positive influence on the COF for all alloys. For the engine oil on the steel alloys small reductions of the COF in the range of a few 10% were found, but it was pointed out that this is close to the uncertainty of these kinds of tests [43]. Other studies found similar values [44, 45].

In contrast, the engine oil reduced the COF for the Ti-based alloys by more than a factor of two, which also reduced the observable wear and the LIPSS remained visible after the test, as it is displayed in Figure 28.8. As can be seen in the optical micrographs and SEM images, the untreated Ti6Al4V surface is relatively flat and the RSTT causes a wide wear track. During the repeated cycles of the RSTT parts of the surface are deformed, debris is generated and dragged around, leading to the ever-changing coefficient of friction. In contrast, the laser treated surface shows only micrometer-sized wear tracks in the LIPSS, meaning low amount of contact between the test-ball and the surface. The debris can also be caught in the ridges of the undisturbed LIPSS, preventing them increasing contact. This results in a relatively stable COF – after the first few sliding cycles – and a reduction in observable wear. It should be pointed out that similar reasoning hold for surfaces that are initially rough in the µm range, like the X30CrMoN15-1 steel in [42] but surface structures on that scale are discussed below.

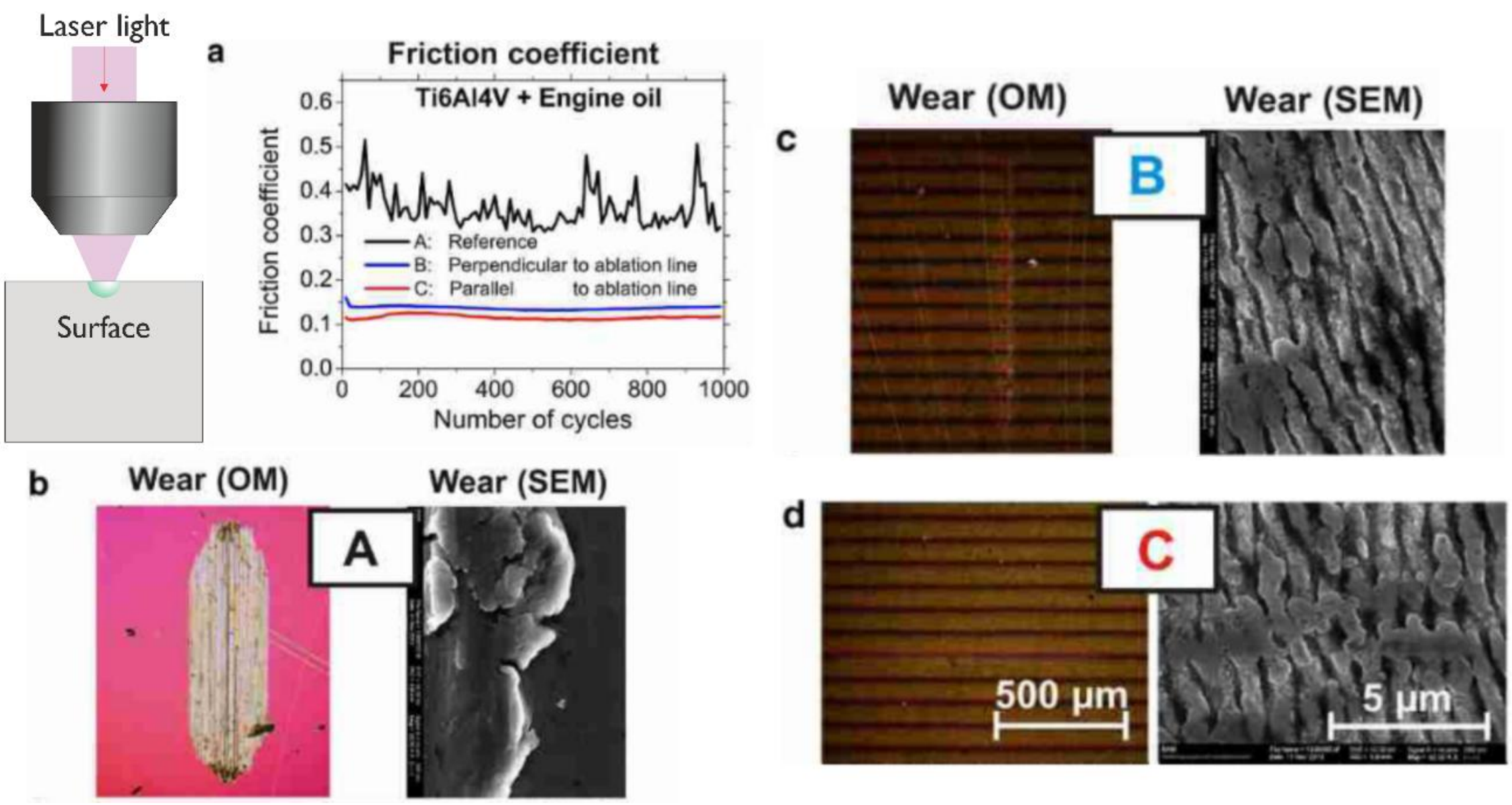

Figure 28.8 (**a**) Plot of the friction coefficient as a function of the number of cycles during reciprocal sliding (normal force 1 N, stroke 1mm, frequency 1 Hz) lubricated with engine oil. (**b**) Optical microscopy (OM) and SEM images of the wear track on a non-laser processed surface. (**c**) Wear tracks across the overall laser scan direction but aligned with the orientation of the LIPSS. (**d**) Wear track along both the overall laser scan direction and the LIPSS. Reprinted (adapted) with permission from Bonse et al. [41], Copyright © 2014 Springer Nature

For the Ti-alloy, this optimized tribological behavior was partially attributed to the change in the chemical properties of the surface by oxidization in combination with additives such as zinc-dialkyl-dithiophosphate (ZDDP) in the engine oil [46, 47] (and references therein). ZDDP forms an anti-wear top-film at the interface of the Ti oxide and the engine oil, preventing the test-ball from making a direct metal-metal contact.

Another contribution to the decrease in friction is the change in surface hardness caused by oxidization. In the case of Ti materials, a decrease in hardness was found by nanoindentation in depths between 100 nm and 200 nm [43, 48]. This coincides with the gradual oxidization of the top layer of the laser nanostructured Ti [38]. The comparison with other oxidization techniques, like heat treatment, determined a lower limit for the oxide layer thickness of ~150 nm for this optimization to work in RSTT [48]. However, an optimization of tribological behavior was also reached by hardening through nitridation [45]. Ti6Al4V alloy was treated by nanosecond laser pulses in nitrogen atmosphere ($\tau$ = 7 ns, $\lambda$ = 1064 nm, $\upsilon$ = 700 kHz, $w_0$ = ~10 µm, $N_{eff} \approx$ 2-9). The COF was reduced by 70%-80% for simultaneous hardening by nitride (measured increase of 50%-95% by nano indentation with loads up to 1000 mN) and the LIPSS surface. After polishing off the LIPSS, leaving only the hardened surface, the COF was still strongly decreased, but less than with LIPSS, showing the influence of both effects in this case.

The surface's resistance to deformation and wear was also argued to be linked to a change in the crystalline structure (phases) close to the surface as displayed in Figure 28.9 ($\tau$ = 290 fs, $\lambda$ = 1064 nm, $\upsilon$ = 700 kHz, $w_0$ = ~20 µm, $N_{eff}$ = 4) [12]. The laser-induced *modification in the bulk* by melting and re-solidification of the Ti6Al4V surface created a top-layer with smaller grain sizes than the untreated volume, as observed in SEM cross-sectional imaging and electron backscatter diffraction (EBSD). Additionally, Pan et al. measured strong compressive residual stress (up to -746 MPa, gradually reaching a depth of 83 µm below the surface) which can hinder crack growth during the strain of the sliding contact in the OCD measurement and therefore reduces wear. In combination with the gradient of oxidization, the surface hardness was improved by a maximum of 16.6% (microhardness at a depth of 15 μm), the OCD reduced by 68.9% and the wear mass loss by 90%, compared to the untreated surface. While the examples listed so far concentrated on metallic alloys, also other materials can benefit. For example, oxidic ceramic composites with metallic Nb, in which LIPSS formed on the Nb flakes and reduced the COF [49].

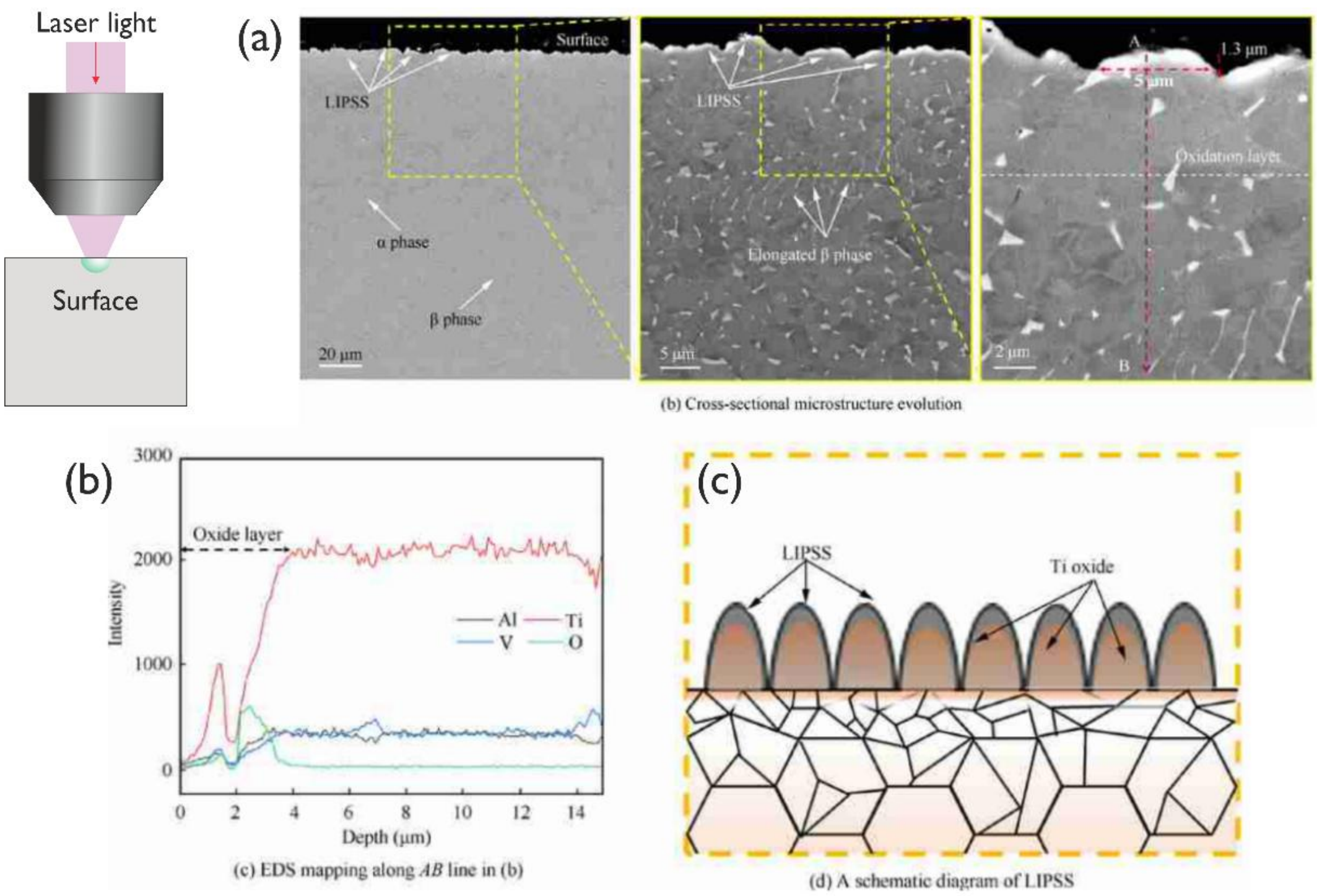

Figure 28.9 (**a**) Cross-section SEM micrographs of laser-irradiated Ti6Al4V titanium alloy with LIPSS on the top surface. The images showcase the microstructure evolution at positions close to the surface and in depth. The alpha and beta phases are present at around 50 and 100 μm, while the surface topography is formed on top of a less than 5 μm thick re-melting coating. (**b**) Energy-dispersive spectroscopy (EDS) elemental mapping on the cross-sections along the vertical red arrow indicated in (**a**), where the extend of the laser induced oxide layer down to 4 μm is highlighted. (**c**) Sketch of the highlighted yellow regions in (**a**) that displays the position of the LIPSS, location of the oxide formation based on the EDS map in (**b**) and the different structural phases in depth. Reprinted (adapted) with permission from Pan et al. [12] under Creative Commons CC-BY 4.0 license, Copyright © 2022 Chinese Journal of Aeronautics, Elsevier

While the presented works to optimize tribological properties with nanoscale surface texturing through LIPSS has motivated research into specialized applications, like investigating the sliding behavior of laser treated Al alloys on snow – as relevant for skiing – [50], larger microscale surface structuring opens another avenue. As presented above, LIPSS and larger laser-induced structures like grooves and spikes can appear at the same time.

In a direct comparison among the self-ordered LIPSS, grooves, and spikes produced by simple laser beam scanning, the LIPSS and grooves presented a similar reduction of COF in RSTT by a factor of 4 – 5 compared to untreated (polished) surfaces, but the spikes exhibited a slightly smaller reduction [38]. SEM investigations showed a strong deformation of the spikes, with the tips of the structures being ground off, which correlates to the COF increasing with the number of cycles until it asymptotes to a higher value than grooves and LIPSS after a few hundred cycles.

Another option for microscale texturing is the individual design of surface morphologies, which can be produced by techniques like Direct Laser Interference Patterning (DLIP), Direct Laser Writing (DLW), Lithography or even imprinting. These can create topographic lines, dimples or other surface shapes. Here, one example for a special application is the cold forging of aluminum, where star- and ring-shaped surface structures were produced by melt pool modulation [51]. The laser-textured surface reduced both the created torque and adhesion of Al, indicating optimized tribological properties.

For such larger surface textures, the discussion about the optimization of the tribological properties reaches into effects on a larger scale than "just" hardness, contact area or surface chemistry. For lubricated contacts, the flow (movement) of the liquid becomes relevant: The 2019 review article by Grützmacher et al. [7] presents a variety of designed surface structures like grooves, grids and dimples that act, for example, as lubricant reservoirs.

Aside from texturing, chemical and structural changes, other novel pulsed laser-assisted surface modifications are conceivable. One example is the additive surface texturing of cutting tools presented by Böhm et al. [15]. In this study, a ms-pulsed laser was used for the implantation of hard ceramic particles into high-speed steel (HSS) cutting tools, contrasting with traditional subtractive methods like laser ablation. The modified tools, featuring dome-shaped ceramic structures on their surface, were tested in dry cutting of ductile cast iron. The results demonstrate a significant reduction in both cutting and feed forces, along with a threefold increase in tool life. These findings highlight the potential of pulsed laser processing to enhance the tribological performance and durability of cutting tools, offering a promising alternative to conventional surface texturing techniques.

In summary, laser-induced surface texturing, particularly through LIPSS, has shown significant potential for optimizing tribological performance by reducing friction and wear. The effectiveness of these modifications depends on multiple factors, including the tribological conditions, material composition, lubricant type, surface chemistry, and structural changes induced by laser processing. While nanostructures like LIPSS can enhance performance through controlled oxidation, increased hardness, and reduced contact area, larger microstructures such as grooves and dimples may offer additional benefits, particularly in lubricated environments where they can influence fluid flow. These findings highlight the versatility of laser-based surface engineering in improving the durability and efficiency of mechanical components, paving the way for spatially resolved tribological solutions across various industries, including the automobile sector, energy generation through bearing rotation, and oleophilicity for friction reduction applications [19, 43].

## *3.2 Wettability Control on Industrially Relevant Materials*

Pulsed laser systems, particularly femtosecond and nanosecond lasers, have gained significant attention in surface modification for various applications, including the control of wettability [52]. The unique nature of laser-induced surface structuring allows precise control over micro- and nanoscale features, leading to modified surface properties that can enhance or reduce wettability depending on the design. One way to identify such suitable designs is to copy strategies from natural surfaces to create Biomimetic Surfaces [5] (see Chap. 24, Brock et al.).

A critical factor influencing wettability in laser-processed materials is the redeposition of ablated material, in the form of debris or particles, which can significantly affect the surface morphology and, consequently, the wetting behavior. Although one of the most studied techniques for modifying wettability using pulsed lasers is the creation of LIPSS [53], in this section, we focus on those applications that study the effect of debris and debris redeposition, as it can significantly alter the surface structure and its wettability.

Figure 28.10 presents a study that explored the effect of combined morphology, including LIPSS and debris redeposited on top, on the wetting performance. In this study, Florian et al. [10] irradiated commercially available steel substrates with a high repetition rate femtosecond laser system, obtaining LIPSS at different fluences. It was found that at higher repetition rates, the base temperature of the steel substrate increased due to laser pulse heat accumulation caused by successive laser pulses, for repetition rates equal to or larger than 250 kHz. The morphology of the redeposited debris was also characterized for the different repetition rates, demonstrating that its morphology depended on the repetition rate used: low repetition rates below 250 kHz led to dust-like microparticles as debris, whereas for higher repetition rates, debris redeposition occurred in the form of spherical nanoparticles. The morphology difference was associated with molten material ejection, especially at high repetition rates, combined with particle shielding close to the surface of the steel substrate. This redeposited material affected the surface morphology of the laser-irradiated regions, which in turn altered the wettability with water. In particular, the study highlighted how controlling laser parameters, such as the repetition rate and pulse energy, was critical to minimizing the unwanted debris deposition and achieving desired wettability outcomes.

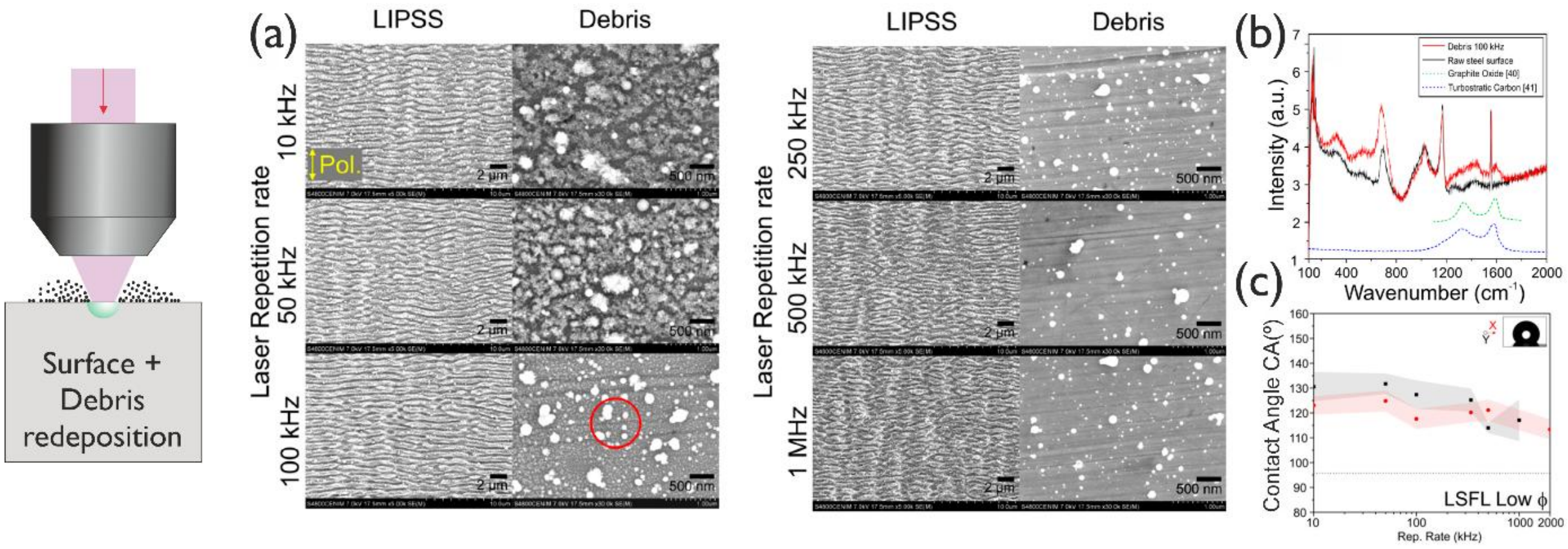

Figure 28.10 (**a**) Top-view SEM micrographs of steel substrates after irradiation with femtosecond laser pulses at constant fluence and number of effective pulses per unit spot. The parameter varied here is the laser repetition rate. The surface structures look similar, but the redeposited debris in the surrounding areas changes from microdust particles to spherical nanospheres. (**b**) The chemical composition of the debris is studied and compared with reported Raman spectra, suggesting the presence of graphite oxide and turbostatic carbon. (**c**) Wetting characterization for the different structures obtained with a 3 µl deionized water droplet. The wetting presents a slight shift from 130º to 110º associated with a variation of morphology and debris redeposition at the irradiated surface. The contact angle was measured after 15 days counting from the laser irradiation day. Reprinted (adapted) with permission from Florian et al. [10] under Creative Commons CC-BY 4.0 license, Copyright © 2022 Materials, MDPI Switzerland

In addition to LIPSS, pulsed laser-induced dewetting of thin metal films is another process where redeposited particles influence wettability. In this process, pulsed laser irradiation causes the metal film to melt and retract due to surface tension, resulting in nanoparticle formation as the molten material re-solidifies. The resulting nanoparticles, which can be redeposited onto the surface, modify its roughness and, consequently, its wettability. A study on pulsed laser-induced dewetting in nanoscopic metal films demonstrated how specific laser parameters could be adjusted to control the size and distribution of the nanoparticles, thus influencing the wettability of the surface [54].

An important aspect of the effect of debris on wettability is that it does not just depend on the morphological change of the surface. It is a known fact that the wetting on laser-induced surface modifications changes over time due to chemical changes of the surface. Wood et al. [55] observed that laser textured stainless-steel surfaces are initially superhydrophilic, and turn superhydrophobic over the span of a few weeks' time due to the formation of a hydrophobic carbonaceous layer, while first storing the sample in boiling water for several days prevented this development. It was shown that there is an interplay between the surface chemistry of the debris particles and the underlying textured surface. Chap. 25 by Peter Gregorčič discusses the importance of surface chemistry and post-laser chemical treatment for wettability of metal surfaces.

A very recent study by Lubig et al. [56] showed that the redeposition of ablated particles influenced the long-term stability of the wettability of stainless steel, with some surfaces exhibiting improved hydrophobicity over time due to the formation of additional debris layers. This study underscores the complex interaction between laser processing, surface debris, and wettability, especially when considering materials in practical applications. An example of the topography of the laser-induced structures acquired via AFM and the corresponding surface's characterization through wetting tests with water is included in Figure 28.11. Importantly, the wetting scenarios (Wenzel or Cassie-Baxter, or combination of both) modify tremendously the chemistry of the deposited debris due to the increased surface contact with the water droplet. Evidence of this is that the wetting characterization conducted on fresh laser irradiated areas displayed behavior significantly different from that performed repeatedly on the same, already measured surface, which suggests that a water-induced oxidation takes place and modifies the wetting behavior of the inspected surfaces.

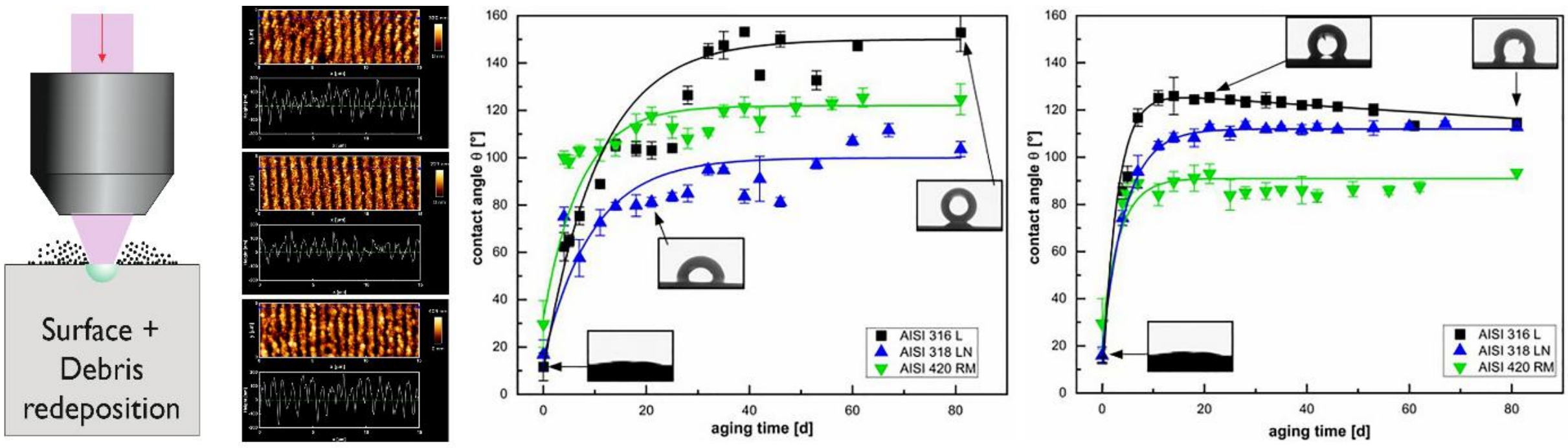

Figure 28.11 On the left: AFM topographical data and profiles taken from three different laser irradiated areas on AISI 316L, AISI 318LN and AISI 420RM steel (from top to bottom). In the middle and on the right: contact angle measurements with water are plotted for the same three surfaces over 80 days on always a fresh part of the surfaces (middle), and on previously measured surfaces (right), displaying insets of the water droplet on selected surfaces/days. Reprinted (adapted) with permission from Lubig et al. [56] under Creative Commons CC-BY-NC-ND license, Copyright © 2022 Applied Surface Science, Elsevier

The impact of debris redeposition on wettability is not restricted to femtosecond lasers. Nanosecond pulsed lasers have also been employed for surface texturing and wettability modification [57]. It has been shown that specific textures induced by debris redeposition play a significant role in determining a laser-processed surface's wetting behavior. These findings underscore the importance of managing debris formation during laser processing to achieve the desired wetting properties. To mitigate the effects of debris redeposition and restore the desired wettability, post-processing cleaning methods are often employed. For example, specific wet chemical treatments could effectively remove redeposited nanoparticles from LIPSS-irradiated areas without damaging the underlying structures [58]. The researchers found that cleaning the debris not only restored the original wettability but also improved the surface's performance in various applications. This highlights the importance of post-processing steps in ensuring that the desired surface characteristics are maintained after laser treatment.

In conclusion, pulsed laser systems offer versatile and powerful methods for modifying surface wettability through techniques such as LIPSS and laser-induced dewetting. However, the phenomenon of debris redeposition during laser processing plays a critical role in determining the final wetting behavior of the surface. This makes it essential to understand the mechanisms of debris formation and redeposition, as well as to control laser parameters to minimize the partly unwanted effects of debris. Additionally, implementing effective post-processing strategies is key to optimizing the wettability of laser-treated surfaces and making the effect long-term stable for practical applications.

## *3.3 Optimized Welding of Materials*

Pulsed laser welding has become an essential technology in modern manufacturing, offering unique advantages in precision and control compared to traditional welding techniques. In this process, a high-energy laser beam is focused onto the material surface, delivering energy in discrete bursts or pulses. This controlled energy input causes localized melting of the material, allowing it to fuse together. While continuous wave (cw) laser welding ensures a steady energy input and the formation of a stable keyhole, pulsed laser beam welding can encounter challenges in maintaining a consistent keyhole, which is essential for deep penetration welding. The rapid variations in pulse energy make it difficult to achieve the continuous penetration typically seen in cw-welding, potentially leading to instability in the weld pool and inconsistent weld quality. This is particularly problematic for applications requiring deep penetration or a smooth, uninterrupted weld seam. However, the advantage of pulsed laser radiation can still provide benefits for many applications. These pulses are typically controlled in terms of duration, frequency, and energy, providing excellent flexibility in managing the heat input and the resulting weld characteristics. The examples range from *surface level modifications* affecting micrometer-thin layers of material to *bulk modifications* melting volumes of hundreds of micrometers.

The pulse duration, which can range from milliseconds to femtoseconds, significantly impacts the laser welding process. Shorter pulses offer high precision and minimize the heat-affected zone (HAZ), which is critical when welding delicate or heat-sensitive materials. The amount of energy delivered per pulse is also critical to achieving the desired penetration depth and weld quality. By adjusting the frequency of pulses, it is possible to tailor the thermal cycles to suit different materials and joint configurations. The controlled, localized heating minimizes the risk of distortion and warping, making pulsed laser welding ideal for applications that require high accuracy and minimal thermal impact. This technique is particularly valuable in industries such as aerospace, automotive, electronics, and medical device manufacturing, where precision and material integrity are paramount. One of the main advantages of pulsed laser welding is its ability to weld dissimilar materials [59–61], which is challenging for many conventional welding methods. The high degree of control over the pulse energy deposition and the ability to focus the beam to a fine point make it possible to create high-quality welds even in thin, highly reflective, or transparent materials [62]. Additionally, the process can be easily automated and integrated into production lines, offering both, high throughput and good repeatability.

Building on these advantages, femtosecond laser pulses extend the capabilities of pulsed laser welding to even more challenging applications, such as joining sapphire, a hard and brittle material, to metal. The ultra-short pulse duration of femtosecond lasers allows for precision welding with minimal heat damage to the sapphire, showcasing the power of pulsed laser welding for microscale applications where traditional welding methods would be impractical [63]. In Figure 28.12, femtosecond laser pulses were employed at different repetition rates and scanning speeds at the interface between the surface of steel and sapphire, causing the materials to weld. The microstructure formed at the interface was studied in detail, discovering that within an interaction layer of 1 µm, both materials mix, producing a stable and strong welding region, without any voids or microcracks. The shear strength of the joints reached 108.35 MPa, 80% higher than that reported for diffusion-bonded or brazed joints for similar materials.

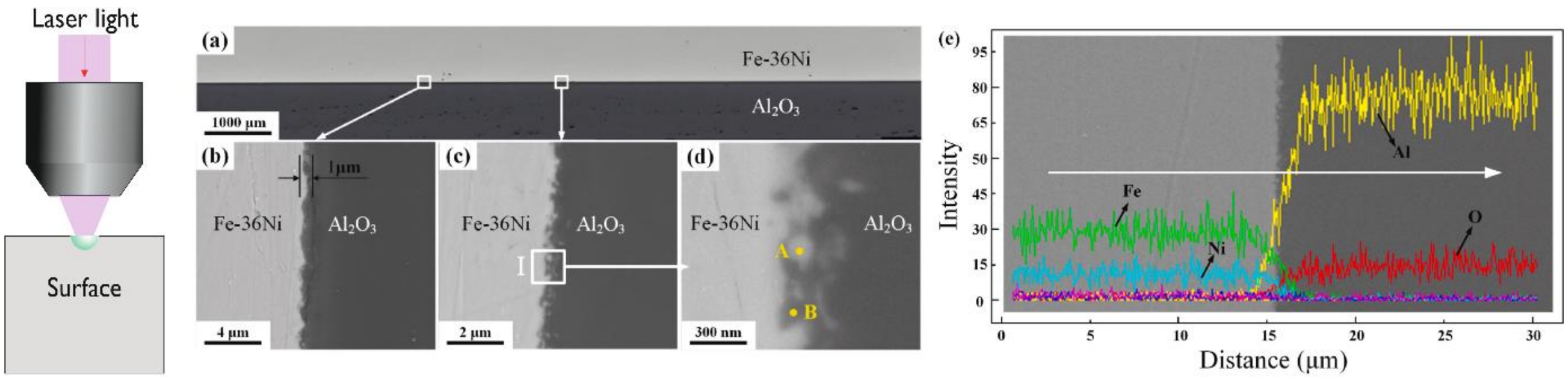

Figure 28.12 (**a**) Cross section of a typical interface between sapphire $Al_2O_3$ and Fe-36Ni welded by femtosecond laser pulses. (**b**, **c**) Magnified and rotated region of the interface allows for determining an interface region of 1 µm. (**d**) At the interface, the white nanoparticles marked by the yellow point **A** correspond to Fe-36Ni alloy and the black matrix marked by the yellow point **B** to sapphire. (**e**) EDS elemental line scan across the interface along the white arrow marked in the figure, displaying relative concentrations of Fe, Ni, Al and O. Reprinted (adapted) with permission from Pan et al. [63], Copyright © 2023 Ceramics International, Elsevier

A recent example with semiconductor materials focuses on the feasibility of using nanosecond laser pulses to weld silicon (Si) and gallium arsenide (GaAs). The research team successfully achieved transmission laser welding between Si workpieces, attaining a shear joining strength of $32 \pm 10$ MPa, which is comparable to existing complex processes. Furthermore, the study extended to welding dissimilar semiconductor combinations, demonstrating the versatility of this technique. These findings suggest that transmission laser welding could serve as a high-efficiency, flexible process beneficial for developing technologies such as lab-on-a-chip and hybrid semiconductor systems [64]. On the other hand, studies on aluminum alloy welding have shown that adjusting laser pulse frequency and laser-wire distance can significantly influence arc shape, metal transfer behavior, and weld formation in hybrid welding processes. Optimal parameters can lead to improved weld quality and efficiency. These studies underscore the adaptability of pulsed laser welding in various industrial applications, highlighting its potential for process optimization and quality enhancement [65].

Pulsed laser radiation can also be used to create complex, dissimilar welds. The studies by Wiegand et al. demonstrated that by using laser radiation pulsed in the ms-range and employing a niobium interlayer, a high-strength joint between a NiTi shape memory alloy and titanium could be achieved [66]. From the same research group, the use of pulsed laser welding to join dissimilar materials, specifically Ti and stainless steel, with biocompatible Nb as a filler material also proved feasible, and it is included in Figure 28.13. The results showed that Nb acted as an effective intermediary in creating strong and durable welds between the two metals, addressing the challenges of joining materials with significantly different thermal properties. Based on the weld seam surfaces, the individual laser pulses that welded Ti and stainless steel with an interlayer of Nb can be seen. The peak/average power studied in this case for material melting started at 300 W and increased until 400 W, where visible cracks appeared, compromising the weld metal structure. The choice of Nb was strategic, aiming to mitigate issues like brittle intermetallic compound formation at the weld interface. Given the high mechanical resilience of the weld seams, combined with the biocompatibility of the selected filler material, the findings of this study establish a fundamental basis for the implementation of this material combination in advanced medical technology applications. The experimental results highlight the potential of pulsed laser welding in fabricating reliable and robust dissimilar metal joints [67].

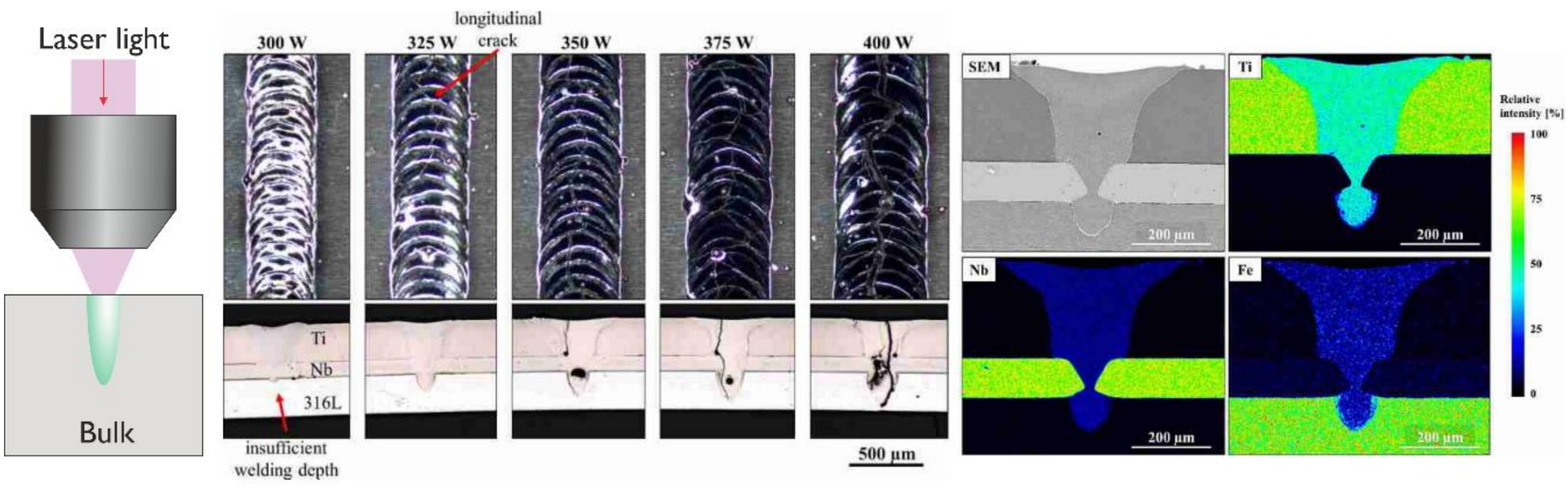

Figure 28.13 On the left: Top views of weld seam surfaces and cross-sections of weld seams generated with a pulse duration of 7 ms and indicated pulse peak powers. Note that the differing colors of the weld seam are attributed to deviations in light reflections resulting from the varying topography of the weld. Visible crack formation occurs for powers higher than 300 W, indicated by the superimposed red arrow. On the right: SEM micrographs of the cross section and corresponding EDS elemental mappings of a crack-free representative sample after adjustment of parameters (pulse duration 10 ms, pulse peak power 300 W) visualizing the relative intensity distribution of Ti, Nb, and Fe in the region of the weld metal. Reprinted (adapted) with permission from Wiegand et al. [67], Copyright © 2023 Metallurgical and Materials Transactions A, Elsevier

These examples illustrate the growing range of applications for pulsed laser welding, from micro-welding of fragile materials like sapphire to joining high-performance alloys in the aerospace and medical industries. The ability to precisely control heat input, minimize material distortion, and weld dissimilar materials has made pulsed laser welding an invaluable tool in advanced manufacturing processes. In extension to the nano- und microscale presented here, Chap. 33, Olschok et al, discusses the question “How Thick Can We Go?” with laser welding in the millimeter to cm range. Also, Chap. 32, Teixeira et al., presents the American Footprint of Laser Processing, including market trends for Laser Beam Welding.

## 3.4 Micro- and Nano-texturing for Adhesion Applications

Laser surface texturing is an advanced technique used to modify the surface properties of materials to improve adhesion in various applications [68, 69]. This process involves directing high-intensity pulsed lasers onto a material surface, resulting in processes including localized heating, melting, ablation, and phase transitions. By adjusting parameters such as pulse duration, energy, and repetition rate, micro- and nanoscale features can be created on the surface. These surface features can range from micro-grooves to nanoscale pits and particles, altering the material roughness, wettability, and chemical composition [70]. These changes can significantly improve surface adhesion by enhancing mechanical interlocking, increasing surface area, or modifying the surface energy of the material, all of which are essential for strong and reliable bonding in different industrial applications [71]. The interaction between the laser and the material is fundamental to achieving the desired surface modifications. Pulsed lasers deliver energy in brief bursts, allowing precise control over the surface modifications without significantly affecting the bulk material. This ability to selectively modify the surface makes laser texturing highly versatile for a range of applications, including in industries where adhesion is critical, such as automotive, aerospace, medical, and manufacturing sectors. In the particular case of biofilm adhesion, there are numerous studies compiled in the following review papers, as the biological species relation with surfaces depends on a wide range of parameters that are not only limited to the morphology and chemistry at the surface, but also on the biological nature of the living species [72–74]. For more medical applications, pulsed laser systems are used to enhance the adhesion of bonding agents to materials like metals, ceramics, and composites [75, 76].

To optimize the mechanical properties of adhesively bonded joints, the creation of micro- or nanostructured surfaces is crucial because these textured surface structures increase the surface area and promote stronger and more durable bonds. This approach is effective when bonding challenging materials such as titanium [77] and stainless steel [14]. The micro- and nanoporous structures formed by debris formation are particularly advantageous for aging-resistant adhesive bonds, as they enable the adhesive to infiltrate and thus provide an increased contact area, as can be seen in Figure 28.14. The increased bond strength additionally results from their mechanical interlocking on the micro- to nanoscale and the formation of a nanocomposite between the adhesive and the metal oxide. Together, these factors improve the adhesive interface, which increases both the adhesive strength and the aging resistance.

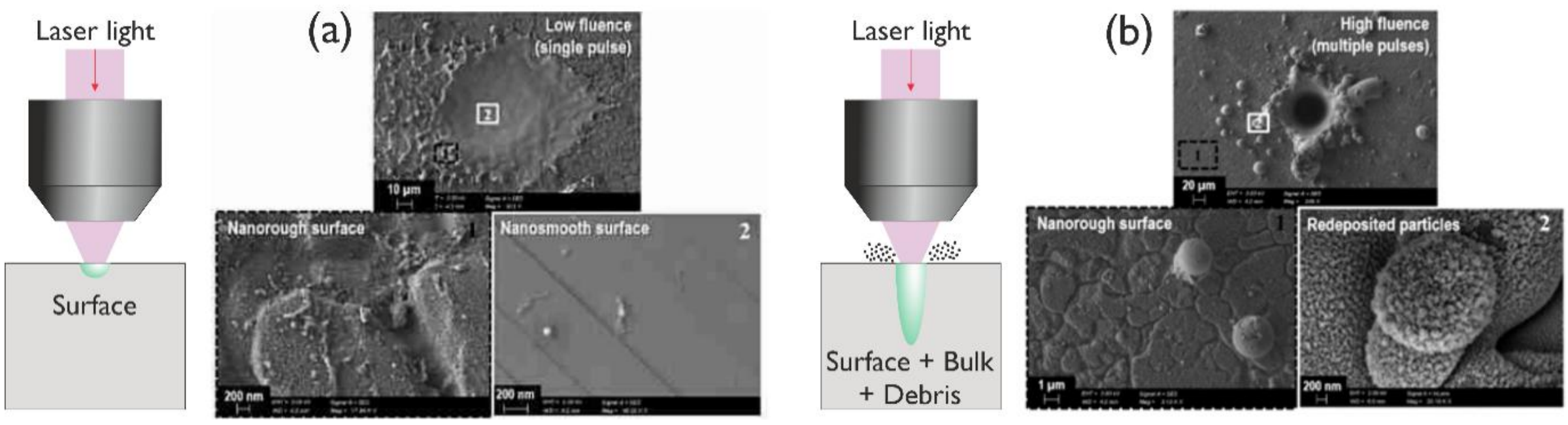

Figure 28.14 (**a**) SEM micrographs of an area irradiated with a single pulse at low fluence. The *surface modification* around and at the laser spot shows nanorough and nanosmooth surfaces, respectively. (**b**) With multiple pulses at higher fluences, the combined *surface and bulk modification* results in a rough topography that includes micrometer sized particles adorned with powder-like nanoparticles. Reprinted with permission from Mechtold et al. [14], Copyright © 2024 Sage Publications

Another study examined the effect of different laser systems on the adhesion of zirconia ceramic surfaces, which are commonly used in dental and medical applications [78]. The surfaces were treated with femtosecond lasers (1027 nm wavelength), carbon dioxide lasers (10,600 nm wavelength), and also treated abrasively by micro blasting them with aluminum-oxide particles [76]. The femtosecond laser, with its extremely short pulse duration, was found to be the most effective in enhancing the adhesion forces on the zirconia ceramic. The laser treatment produced micro- and nanoscale surface structures, as shown in Figure 28.15, which increased the surface area and roughness, thereby improving the mechanical interlocking between the adhesive and the substrate. In comparison, $CO_2$ lasers, while effective, did not achieve the same level of precision, and aluminum-oxide particle treatment did not result in adhesion as strong. The study highlighted the advantages of femtosecond lasers in achieving superior adhesion strength, making them highly suitable for applications requiring strong bonding, such as dental restorations.

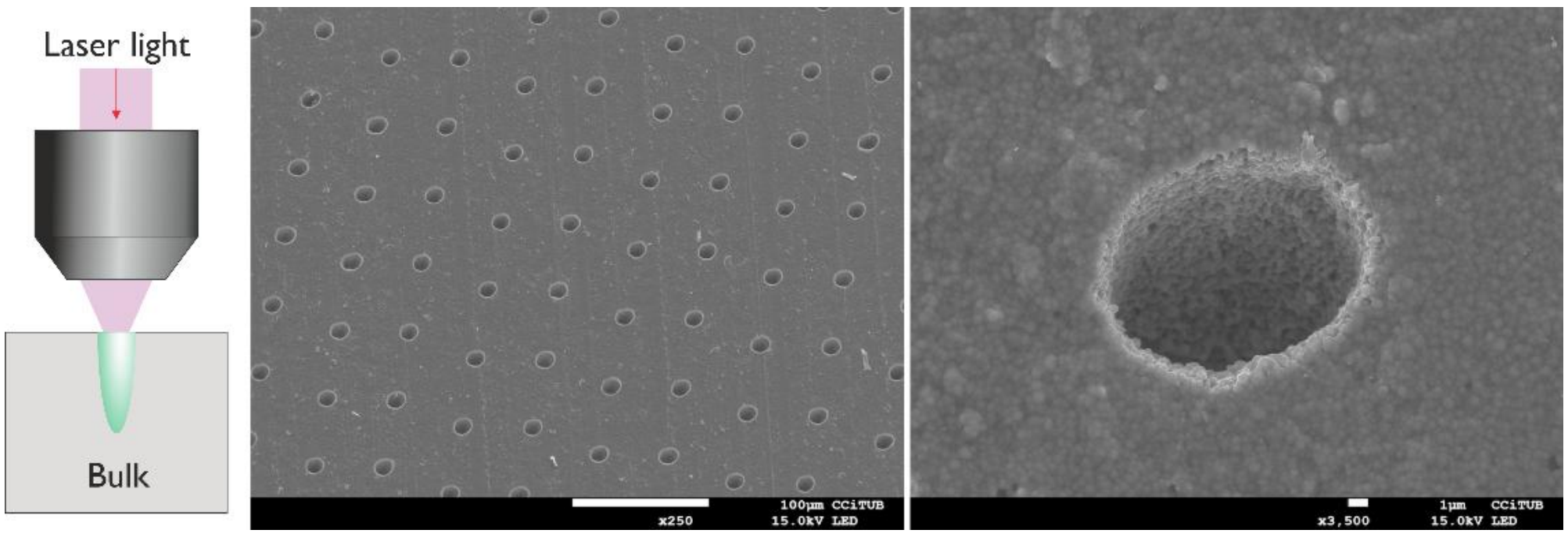

Figure 28.15 Femtosecond irradiation on zirconia substrate, producing an array of micro-holes designed to be used as pinning points for glue applied in dental prosthesis. The *bulk modification* causes minimal thermal damage to the surrounding laser irradiated area. A porous non-periodic structure is visible inside the craters as the SEM images were acquired at an angle. Reprinted (adapted) with permission from Piulachs et al. [76], Copyright © 2023, Lasers in Medical Science, Springer Nature

Femtosecond laser treatment has also been employed to tailor resin-metal adhesion, which is relevant in many industrial applications. In one study by Rathnayaka and Kietzig [71], femtosecond lasers were used to create hexagonal surface textures by raster scanning on titanium and aluminum alloys to alter the bonding with resin materials compared to non-textured surfaces, as it is displayed in Figure 28.16. The laser treatment modified the surface at both micro- and nanoscales, improving the mechanical interlock between the resin and metal. At the same time, the hexagonal holes caused bubble trapping at some locations, which promoted crack propagation due to residual stress build-up. While the overall tensile strength increased, because the resin still interacted with the nanostructures at the hole entrance, this design opens a pathway to control and possibly even reduce the adhesion.

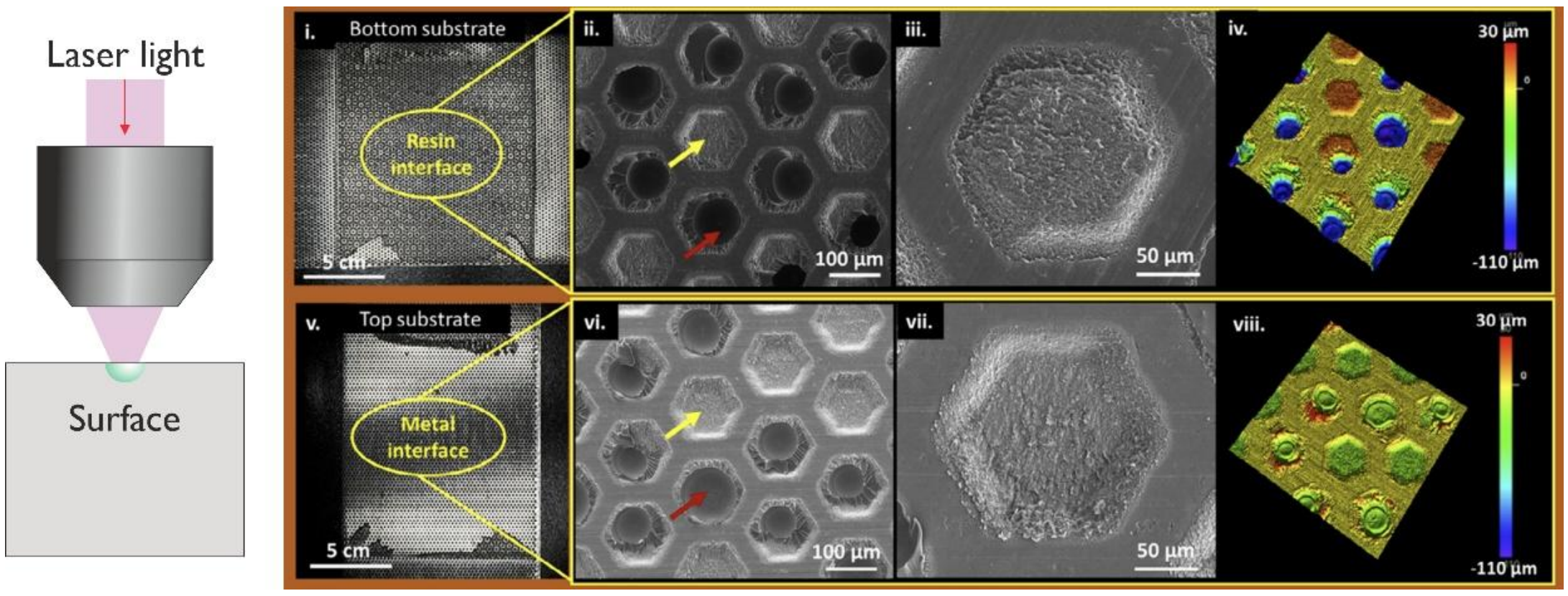

Figure 28.16 SEM images and 3D confocal images of the laser structured metal (bottom row) and resin (top row) interface after failure in tensile testing. Yellow arrows in **ii** and **vi** indicate regions of adhesive failing, where the LIPSS texture is imprinted in the resin (details in **iii** and **vii**). Red arrows indicate voids due to bubble trapping during application and/or curing of the resin. Reprinted with permission from Rathnayaka and Kietzig [71] under Creative Commons CC-BY-NC-ND license, Copyright © 2022 Physica Status Solidi (A), by permission from John Wiley and Sons

Overall, pulsed laser surface texturing is a highly effective technique for enhancing adhesion properties across a wide range of materials. By enabling precise control over surface morphology at the micro- and nanoscale, lasers provide a versatile and efficient means of improving mechanical interlocking, surface energy, and chemical bonding, making them invaluable for various industrial and medical applications. The continued exploration of laser parameters and their effects on adhesion strength will further advance the potential of this technology to improve the performance and reliability of adhesive joints.

# 4 Summary and Outlook

The advancements in pulsed laser surface treatments have paved the way for precise control over material properties at both micro- and nanoscales. This chapter presented how the laser-induced modification of a surface – topologically, chemically, and structurally – on different scales can optimize its properties, such as hardening for reduced wear in tribological applications, or completely alter them toward new applications, such as coloring by diffraction and controlled wettability. Part III of this book will present a variety of other applications ranging from “minuscule scales” like Data Storage (Chap. 42 by Zhang et al.), to “grand scales” like energy, (Chap. 35 by Soldera et al.), aviation (Chap. 39 by Haasler et al.), and space applications (Chap. 40 by Eismann et al.).

As research in this field continues to evolve, several promising directions are expected to shape future developments, particularly in the scalability and efficiency of laser processing techniques for industrial applications. One of the key areas of interest is the refinement of inter- and intra-pulse feedback mechanisms, as summarized here and discussed in more detail in Part 1 of this book. By further improving real-time monitoring and adaptive control, researchers can achieve more precise modifications tailored to specific industrial needs (see Chap. 20 by Bonse et al.). The ability to dynamically adjust laser parameters based on in-situ measurements will further improve the reproducibility and efficiency of surface structuring techniques. Additionally, advancements in ultrafast laser technology, particularly in high-power and high-repetition-rate femtosecond lasers (see Chap. 2 by Gailevičius et al.), will provide enhanced control over energy deposition, leading to more uniform and controlled surface modifications. These developments will play a critical role in improving the precision and consistency of laser-based surface engineering.

The distinction between nano- and microscale modifications will continue to play a crucial role in tailoring surface functionalities. A deeper understanding of how different laser parameters influence the formation of specific structures will enable optimized texturing processes, which in turn can lead to superior material performance in applications such as adhesion, tribology, and wettability control. Future work will also explore the combination of surface topographical modifications with functional coatings to create hybrid surfaces with enhanced mechanical and chemical properties. The integration of advanced computational models and machine learning algorithms will further aid in predicting and optimizing these processes, making them more efficient and reproducible across different material systems, as presented in Chap. 21 by Zwahr et al.

In addition to topographical changes, the chemical and structural modifications induced by pulsed lasers open new opportunities for advanced material design. Future research will likely focus on fine-tuning these changes to develop multifunctional surfaces with enhanced durability, corrosion resistance, and biocompatibility. This will be particularly relevant in fields such as biomedical engineering, aerospace, and energy storage. Furthermore, the controlled introduction of tailored surface chemistries through pulsed laser treatment could lead to breakthroughs in bio-inspired surface functionalities (see also Chap. 24, Brock et al.), such as self-cleaning, antibacterial, or anti-icing properties. This will require an in-depth understanding of laser-material interactions at the atomic and molecular levels, as well as the long-term stability of such modifications under real-world operating conditions.

From an application perspective, the integration of laser-based surface modifications in large-scale manufacturing remains an ongoing challenge. Several chapters in this book tackle the current progress in this field. In Chap. 8 of Part I, "Fundamentals", Gedvilas et al. provide a general overview of the move of ultrashort pulse lasers "from Science to Industry," including relevant modeling and processing strategies. Part II of this book focuses on how to scale up laser processing. Among many chapters discussing more specialized topics, Angurel et al. in Chap. 16 comprehensively present strategies and challenges for transferring laser processing from the laboratory to production and Schille et al. in Chap. 22 detail how new technologies and strategies lead to increasingly larger processing speeds. Lastly, we would like to point to Chap. 34 by Hennig at al. which introduces the production of nano- and microscale structures by embossing, where the pulsed laser might not touch the final product but is the essential technology for creating the master structures. All of these, and the many other reports in this book, prove the recent progress in scaling and applying laser processing., from the advancements in laser technology, which provide more powerful pulsed laser sources, to inventive laser processing strategies, like burst-processing, or large-scale production methods such as embossing.

Future advancements in laser processing, such as coupling with artificial intelligence-driven optimization, hold the potential to facilitate industrial adoption. While high-throughput manufacturing could benefit from the implementation of high-power, high-repetition-rate femtosecond laser sources, challenges remain in achieving the same fine precision at the nano- and microscale as slower scanning methods. Similarly, the integration of high-speed scanning systems, like polygonal mirrors and galvanometric scanners, could enhance processing speeds, but their application to non-planar surfaces and 3D geometries remains a significant industrial challenge. Future developments, including the recent progress presented in this book, will become critical in industries such as automotive, aerospace, and biomedical sectors, where large-scale yet precise modifications are required. Additionally, the combination of laser texturing with roll-to-roll processing techniques could unlock new possibilities for cost-effective, scalable production of functional surfaces.

Overall, pulsed laser surface treatments hold immense potential for revolutionizing material processing across various sectors. By continuing to refine the underlying mechanisms, improving process efficiency, and exploring novel applications, this technology will remain at the forefront of next-generation material engineering solutions. Chap. 31 by Cheng at al. and Chap. 32 by Teixeira et al. give an overview of the recent and current efforts of this in the Asian and European market, including a focus on the contributions to more sustainable and a circular manufacturing. The future of pulsed laser treatments will be shaped by interdisciplinary collaborations between academia and industry, enabling the transition from laboratory-scale research to fully industrialized, high-throughput solutions. With continuous advancements in laser technology, real-time process control, and scalable implementation strategies, pulsed laser processing is poised to become a cornerstone of modern material engineering and manufacturing.